\documentclass[aps, prb, twocolumn, superscriptaddress, longbibliography]{revtex4}
\usepackage{amssymb}
\usepackage{amsmath}
\usepackage{eurosym}
\usepackage{graphicx}
\usepackage{epstopdf}
\usepackage{dcolumn}
\usepackage{bm}
\usepackage{here}
\usepackage[utf8]{inputenc}
\usepackage[T1]{fontenc}
\usepackage{mathptmx}
\usepackage{hyperref}

\begin{document}

\title{Oxygen vacancies in BaTiO$_{3}$ based ferroelectrics: electron
doping, history dependence of $T_{\text{C}}$ and domain wall pinning}

\author{Francesco Cordero}
\email{francesco.cordero@ism.cnr.it}
\affiliation{Istituto di Struttura della Materia-CNR (ISM-CNR), Area della Ricerca di
Roma - Tor Vergata, Via del Fosso del Cavaliere 100, I-00133 Roma, Italy}
\author{Floriana Craciun}
\affiliation{Istituto di Struttura della Materia-CNR (ISM-CNR), Area della Ricerca di
Roma - Tor Vergata, Via del Fosso del Cavaliere 100, I-00133 Roma, Italy}
\author{Paulo Sergio da Silva, Jr.}
\affiliation{Department of Physics, Federal University of S\~{A}\pounds o Carlos,
13565-905 S\~{a}o Carlos, Brazil}
\author{Michel Venet Zambrano}
\affiliation{Department of Physics, Federal University of S\~{A}\pounds o Carlos,
13565-905 S\~{a}o Carlos, Brazil}
\author{Elisa Mercadelli}
\email{elisa.mercadelli@issmc.cnr.it}
\affiliation{Institute of Science, Technology and Sustainability for Ceramics - CNR
(ISSMC-CNR), Via Granarolo 64, 48018 Faenza, Italy}
\author{Pietro Galizia}
\affiliation{Institute of Science, Technology and Sustainability for Ceramics - CNR
(ISSMC-CNR), Via Granarolo 64, 48018 Faenza, Italy}
\date{\today }

\begin{abstract}
We measured the complex Young's modulus of BaTiO$_{3-\delta }$, (BT) Ba$_{x}$%
Sr$_{1-x}$TiO$_{3-\delta }$ (BST) and (Ba$_{0.85}$Ca$_{0.15}$)(Zr$_{0.1}$Ti$%
_{0.9}$)O$_{3-\delta }$ (BCTZ) during heating and cooling runs at various O
deficiencies and aging times. The elastic energy loss has peaks due to the
jumps of isolated O vacancies (V$_{\text{O}}$) and reorientations of pairs
of V$_{\text{O}}$ in the paraelectric phase, from which the respective rates
and activation energies are measured. These rates control the mechanisms of
domain clamping, pinning, fatigue, and anything related to the V$_{\text{O}}$
mobility. In the ferroelectric (FE) phase, the drop of the losses due to the
domain wall motion upon introduction of V$_{\text{O}}$ monitors the degree
of pinning. In addition, large shifts of $T_{\text{C}}$ are observed at the
same value of $\delta $ upon varying the permanence time in the FE state, up
to $\Delta T_{\text{C}}=$ 21~K in BST, while no aging effect is found in
BCTZ.

The phenomenology is explained by considering that $T_{\text{C}}$ is
depressed mainly by the mobile electrons doped by V$_{\text{O}}$. Each
isolated V$_{\text{O}}$ dopes two electrons as itinerant Ti$^{3+}$ ions,
but, when it forms a stable linear V$_{\text{O}}$--Ti$^{2+}$--V$_{\text{O}}$
pair, the two electrons of the Ti$^{2+}$ are subtracted from the mobile
ones, halving doping. The rise of $T_{\text{C}}$ during the initial aging is
then explained in terms of the progressive aggregation of the V$_{\text{O}}$%
. Prolonging aging for years leads to a decrease of $T_{\text{C}}$,
explained assuming that the most stable position of a V$_{\text{O}}$ is at $%
90^{\circ }$ domain walls, whose geometry is incompatible with the pairs.
Then, after enough time the initially aggregated V$_{\text{O}}$ within the
domains dissociate to decorate the $90^{\circ }$ walls, increasing doping
and lowering $T_{\text{C}}.$

The absence of such effects in BCTZ is due to larger activation energy for
pair reorientation and pair binding energy. Then, at room temperature
practically all V$_{\text{O}}$ are paired and static over a time scale of
hundreds of years, explaining the superior resistance of BCTZ to fatigue.
\end{abstract}

\maketitle


\section{Introduction}

The contribution of O vacancies (V$_{\text{O}}$) to fatigue and in general
to the degradation of the properties of perovskite ferroelectrics has been
studying since many decades \cite{GGH15} and is far from being fully
understood. Recently it has even been proposed that many observations
generally ascribed to mobile V$_{\text{O}}$ on the basis of a rather loose
estimation of V$_{\text{O}}$ hopping barrier of $\sim 1$~eV, are rather due
to electronic mechanisms, \cite{Tyu20} especially after observing that the
activation energy for the mean time to failure of multilayer ceramic
capacitors is $\sim 1.6$~eV, close to a half the energy band gap of BaTiO$%
_{3}$ \cite{CHL24}. Yet, it is undeniable that V$_{\text{O}}$ are present,
especially in devices, such as multi layer ceramic capacitors (MLCCs),
having electrodes made of oxidizable non noble metals. Such electrodes
require that the ceramics are sintered in reducing atmosphere, with a
subsequent mild oxygenation, which does not fully eliminate the V$_{\text{O}%
} $ \cite{OAB03,ZTW24}.

On the other hand, V$_{\text{O}}$ are not only detrimental to the materials
properties, since the traditional method for hardening the piezoelectrics,
namely reducing their losses, through domain wall (DW) pinning is to
introduce V$_{\text{O}}$ by acceptor doping \cite{GGH15}. Another possible
effect of introducing acceptor-V$_{\text{O}}$ pairs is enhancing the
electrostrain in piezoceramics, though the exact mechanisms involved are not
yet clear \cite{HWZ24}, but all these effect are greatly affected by rises
in temperature, due the increased mobility of V$_{\text{O}}$ \cite{ZSQ22}.

It is therefore important to improve our knowledge of the behavior of V$_{%
\text{O}}$ in ferroelectric materials at the atomic scale: their mobility,
how they aggregate, and are trapped by dopants and DWs. This is usually done
with methods that probe the long range diffusion, such as measuring the
change of mass or resistivity of samples during annealing in controlled
atmospheres, or the isotope tracer diffusion technique \cite%
{OAB03,DeS15,CSS81b}. These methods, however, reveal the overall diffusion
and not the different steps involved, especially because carried out at high
temperatures, where the aggregation of V$_{\text{O}}$ into pairs and chains
\cite{Cor07,ECC17} becomes irrelevant.

Indeed, selectively probing the different types of jumps of V$_{\text{O}}$
and measuring their concentration is not easy. An isolated V$_{\text{O}}$
has an anisotropic elastic dipole but no electric dipole; therefore its
hopping cannot be probed by dielectric but only by anelastic spectroscopy,
unless it is associated with another charged point defect, \textit{e.g.}
cation vacancy, to form an electric dipole. Also NMR relaxation can
selectively probe the different types of jumps of V$_{\text{O}}$, but it has
been used only few times \cite{HK91,BTK11}, while EPR has been used only for
studying the kinetics of V$_{\text{O}}$ trapped by acceptors \cite{Eic07}.
As a consequence, the link between the V$_{\text{O}}$ mobility and fatigue
or degradation phenomena in ferroelectric perovskites is generally vague,
without determining a detailed microscopic mechanism. Certainly the ionic
contribution of V$_{\text{O}}$ to conductivity is negligible compared to the
polaronic and electronic ones \cite{TS20}, but their hopping as reorienting
elastic dipoles, as electric dipoles when paired with a charged defect, and
their migration to domain walls differently affect the ferroelectric
properties at room temperature over time scales from minutes to months or
years, even in the absence of electric excitation.

Here we present a study of the mobility and clustering of V$_{\text{O}}$
introduced by reducing treatments in BaTiO$_{3}$ (BT), Ba$_{x}$Sr$_{1-x}$TiO$%
_{3}$ (BST) and (Ba$_{0.85}$Ca$_{0.15}$)(Zr$_{0.1}$Ti$_{0.9}$)O$_{3}$
(BCTZ), which are found to have very different hopping rates, depending on
their aggregation state and material composition, and introduce history
dependent effects on the Curie temperature $T_{\text{C}}$. Novel microscopic
insight is provided on how V$_{\text{O}}$ contribute to electron doping and
pinning of domain walls, finally leading to aging and fatigue. In
particular, it is shown how the electron doping is strongly affected by the
degree of clustering of the V$_{\text{O}}$.

\section{Experimental}

Two BaTiO$_{3}$ samples BT1 and BT2 were cut from a bar prepared by
conventional mixed-oxide powder technique at the Department of Chemistry of
the Martin Luther University Halle (Saale), Germany, as described in \cite%
{CTC19}. The present sample BT1, with dimensions $43\times 4.1\times 0.59$~mm%
$^{3}$, corresponds to bar No. 2 in that paper, while the present sample BT2
is a bar $43\times 4.1\times 1.1$~mm$^{3}$. Sample BT3 was another bar $%
32.9\times 4.2\times 1.15$~mm$^{3}$ of BaTiO$_{3}$ prepared in the same
laboratory. Sample BT4, $42\times 6.3\times 0.67$~mm$^{3}$, was prepared,
again by conventional mixed-oxide powder technique, at the Physics
Department of UFSCar, S\~{a}o Carlos (Brazil), as described in \cite{CTC19},
and corresponds to sample No. 1.1 of that paper and sample BT \#1 in \cite%
{CTQ21}.

The BST samples were prepared in UFSCar, as described in \cite{CTS23}. The
bar with $x=$ 0.03 had dimensions $34.4\times 4.4\times 0.43$~mm$^{3}$.

The (Ba$_{0.85}$Ca$_{0.15}$)(Zr$_{0.1}$Ti$_{0.9}$)O$_{3}$ was synthesized
via the solid-state reaction method using BaCO$_{3}$ (Aldrich, $>99\%$
purity), CaCO$_{3}$ (MERCK, $>99\%$ purity), TiO$_{2}$ (Degussa P25, $%
>99.5\% $ purity), and ZrO$_{2}$ (MEL, SC 101, $>99\%$ purity) as starting
materials. The oxide powders were mixed and ball--milled for 20 hours, then
pressed and calcined at 1300~${^{\circ }}$C for 5 hours. The calcined
powders were further ball--milled for 96 hours and sieved to 200~$\mu $m.
Green bars were formed by uniaxial pressing at 100~MPa, followed by cold
isostatic pressing at 300~MPa. Sintering was performed in a covered ZrO$_{2}$
box with samples embedded in BCTZ-50 powders, at 1500~${^{\circ }}$C for 4
hours, with natural cooling in the furnace. Two bars were cut to $47.9\times
5.5\times 0.57$~mm$^{3}$ (BCTZ1) and $33\times 5.5\times 0.55$~mm$^{3}$
(BCTZ2).

Oxygen vacancies were introduced by exposing the bars to a flow of 0.1CO +
0.9Ar at 1~bar and high temperature ($970-1250$~$^{\circ }$C) for up to 2~h,
followed by 1 h homogenization at 800~$^{\circ }$C in the same reducing flow
and cooling to room temperature (RT) in few minutes. The sample was first
wrapped in a Pt foil with open ends, using slabs of Al$_{2}$O$_{3}$ or
Y-stabilized ZrO$_{2}$ to avoid the contact with Pt. The Pt envelope was
placed in a quartz tube with water cooled jacket and heated by induction.
Its temperature was measured with a pyrometer Land M1 600/100C-S and a
hot-wire pyrometer Pirottico SAE, insensitive to changes in emissivity. The
concentration of V$_{\text{O}}$ was estimated from the mass reduction after
the treatment, and might be overestimated by unwanted loss of BaO,
especially at the highest temperatures. In fact, while the total mass loss
of BT1 after all the treatments corresponded to $\delta =$ 0.0118, the mass
gain after reoxygenation corresponded to $-\delta =$ 0.00763, but the sample
was again pale yellow with no trace of V$_{\text{O}}$ in the anelastic
spectrum. Sample BT2 was subjected to the reductions together with BT1 for
control, without measuring its anelastic spectrum. Its state with $\delta
=0.00935$ corresponds to the final $\delta =$ 0.0118 of BT1, and the fact
that $\delta $ estimated from the total mass losses is smaller in the
thicker sample is in agreement with the hypothesis of an overestimation of $%
\delta $ from unwanted losses of BaO.

The complex Young's modulus $E=$ $E^{\prime }+iE^{\prime \prime }$ was
measured by suspending the sample with two thin thermocouple wires in vacuum
and electrostatically exciting its free-free flexural resonances, as
described in \cite{CDC09}. In order to short the thermocouple wires, one of
which grounded, and make the sample conductive in correspondence with the
exciting/measuring electrode, silver paint was applied to the sample. The
Young's modulus is measured from the resonance frequency \cite{NB72}%
\begin{equation}
f=1.028\frac{h}{l^{2}}\sqrt{\frac{E}{\rho }}~,  \label{EqFlex}
\end{equation}%
where $l$, $h$, and $\rho $ are the sample's length, thickness, and density,
assumed as constants, since they vary much less than $E$ with temperature.
The elastic energy loss, $Q^{-1}=$ $E^{\prime \prime }/E^{\prime }$, was
measured from the decay of the free oscillations, after switching off the
excitation, or from the width of the resonance curves. The frequency
dependence of the anelastic spectra was measured by exciting up to three
flexural modes, with frequencies in the ratios $1:5.4:13$, during the same
run. The anelastic relaxation from the hopping of a V$_{\text{O}}$, and
consequent reorientation of its elastic dipole, causes a Debye peak in the
energy loss measured at frequency $\omega /2\pi $, with maximum at the
temperature such that $\omega \tau \simeq 1$, where $\tau $ is close to the
mean hopping time \cite{NB72}.

\section{Results\label{Results}}

\subsection{BaTiO$_{3-\protect\delta }$}

\begin{center}
\begin{figure}[h]
\includegraphics[width=8.5cm]{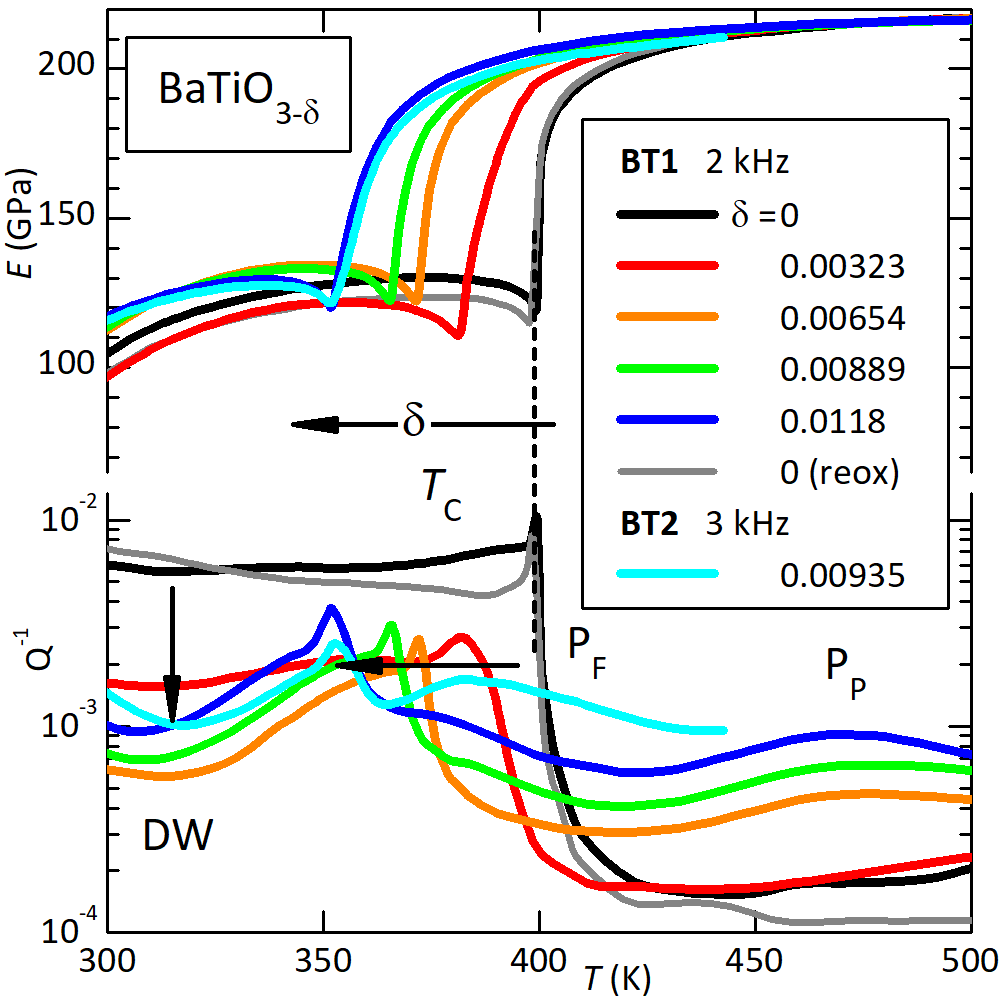}
\caption{Young's modulus $E$ and elastic energy loss $Q^{-1}$ of the two
samples BT1 and BT2 cut from the same bar of BaTiO$_{3}$ at various O
deficiencies $\protect\delta $.}
\label{fig-vs-d}
\end{figure}
\end{center}

Figure \ref{fig-vs-d} presents the anelastic spectra of sample BT1 of BaTiO$%
_{3-\delta }$ at increasing O deficiencies $\delta $. The curves are
perfectly similar to those measured in BT4 and already published \cite{CTQ21}%
. The introduction of V$_{\text{O}}$ has three major effects: \textit{i)}
shift $T_{\text{C}}$ to lower temperature, \textit{ii)} introduce thermally
activated anelastic relaxation peaks in the paraelectric phase, \textit{iii)}
substantially reduce the elastic energy loss from the relaxational motion of
DWs in the tetragonal ferroelectric phase. Peak P$_{\mathrm{F}}$, only
visible when $T_{\text{C}}$ is lowered enough, is due to the hopping of
isolated (free) V$_{\text{O}}$, while peak P$_{\mathrm{P}}$ to the
reorientation of V$_{\text{O}}$ pairs \cite{CTQ21}. At higher temperature,
two minor peaks are found, which grow with $\delta $ much less than P$_{%
\mathrm{F}}$ and P$_{\mathrm{P}}$, and are attributed to V$_{\text{O}}$
trapped by native defects, such as Ba vacancies \cite{CTQ21}.

All these curves are measured during cooling within few days after the
reduction treatment or after a temperature run extending at least to 500~K,
therefore after few days in the FE state. The last curve of BT1 (thin gray)
was obtained after reoxygenating the sample in air at 1100~$^{\circ }$C for
90 min. It is very similar to the initial black curve, and has a
high--temperature background even lower, indicating that no major
degradation occurred during the previous reductions.

The anelastic measurements on BT2, which had been always reduced together
with BT1, started 6 years after the last reduction and the curve in Fig. \ref%
{fig-vs-d} was measured during the fourth cooling cycle. It is very close to
the blue curve of BT1, supposedly in a similar state, except for a higher
intensity of peak P$_{\mathrm{F}}$.

\begin{center}
\begin{figure}[h]
\includegraphics[width=7 cm]{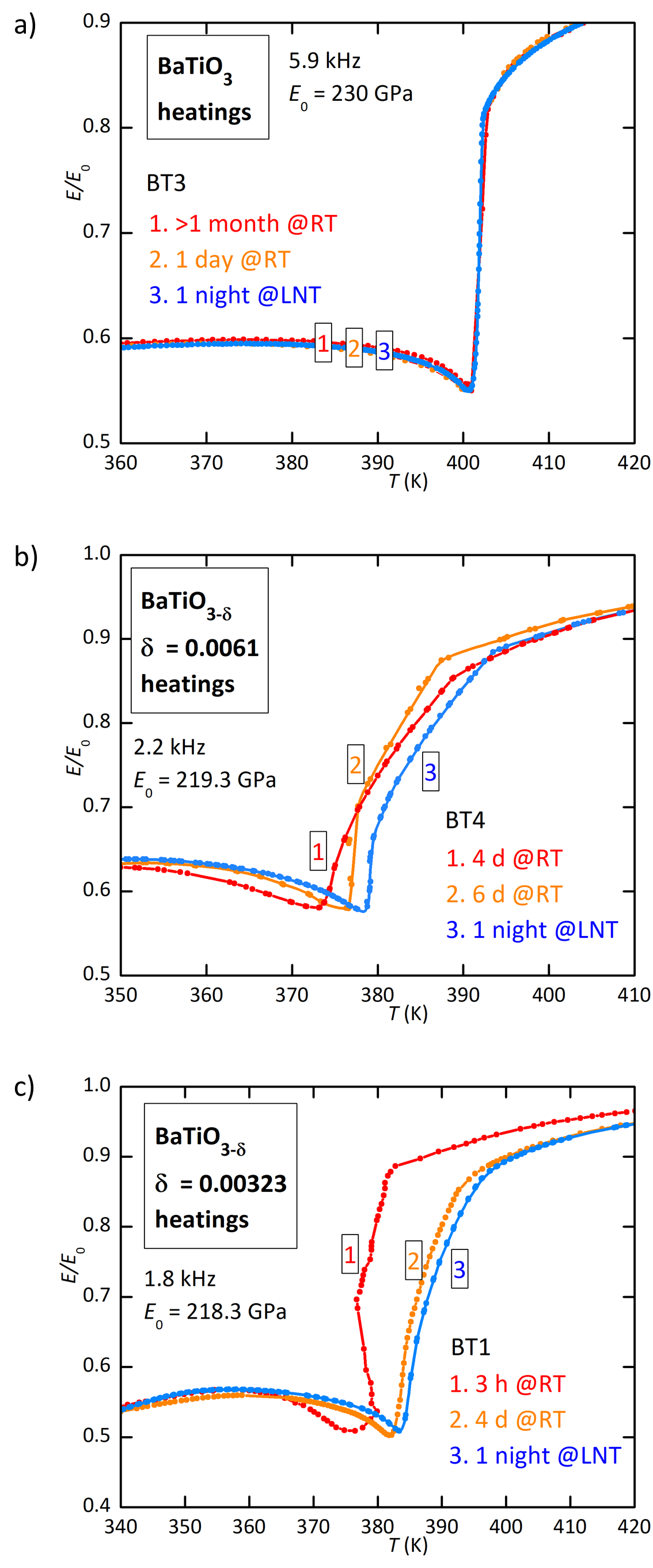}
\caption{Normalized Young's modulus of BaTiO$_{3-\protect\delta }$ measured
around $T_{\text{C}}$ in various conditions. a) sample BT3 with $\protect%
\delta =$ $0$ after $>1$~month (1) and 1 day (2) at room temperature (RT)
and after one night at liquid N$_{2}$ temperature (LNT) (3). b) sample BT4
with $\protect\delta =0.0061$ after 4 days (1) and 6 days (2) at RT and
after one night at LNT (3). c) sample BT1 with $\protect\delta =0.0032$
after 3~h (1) and 4 days (2) at RT and after one night at LNT (3); in curve
(1) there was an inversion of the temperature rate through the transition.}
\label{fig-TcBT}
\end{figure}
\end{center}

Figure \ref{fig-TcBT} presents various $E\left( T\right) $ curves of BaTiO$%
_{3-\delta }$ around $T_{\text{C}}$ measured under different conditions.
Apart from differences in the precursor softening above $T_{\text{C}}$,
which may also be affected by an additional transition above $T_{\text{C}}$
(see beginning of Discussion), the interest is in the fact that $T_{\text{C}%
} $, at the lower edge of the curves, may vary considerably in the same
sample at constant O deficiency. As expected, $T_{\text{C}}$ is stable in
the absence of V$_{\text{O}}$ (Fig. \ref{fig-TcBT}a), but it depends on how
long the sample aged in the ferroelectric phase when measured during heating
in O deficient samples, already with $\delta $ as low as 0.003. In Figs. \ref%
{fig-TcBT}b,c there is an increase of $T_{\text{C}}$ with increasing the
aging time at RT in the range of hours and days, with the highest $T_{\text{C%
}}$ achieved after a night at liquid N$_{2}$ temperature (LNT).

\begin{center}
\begin{figure}[h]
\includegraphics[width=7 cm]{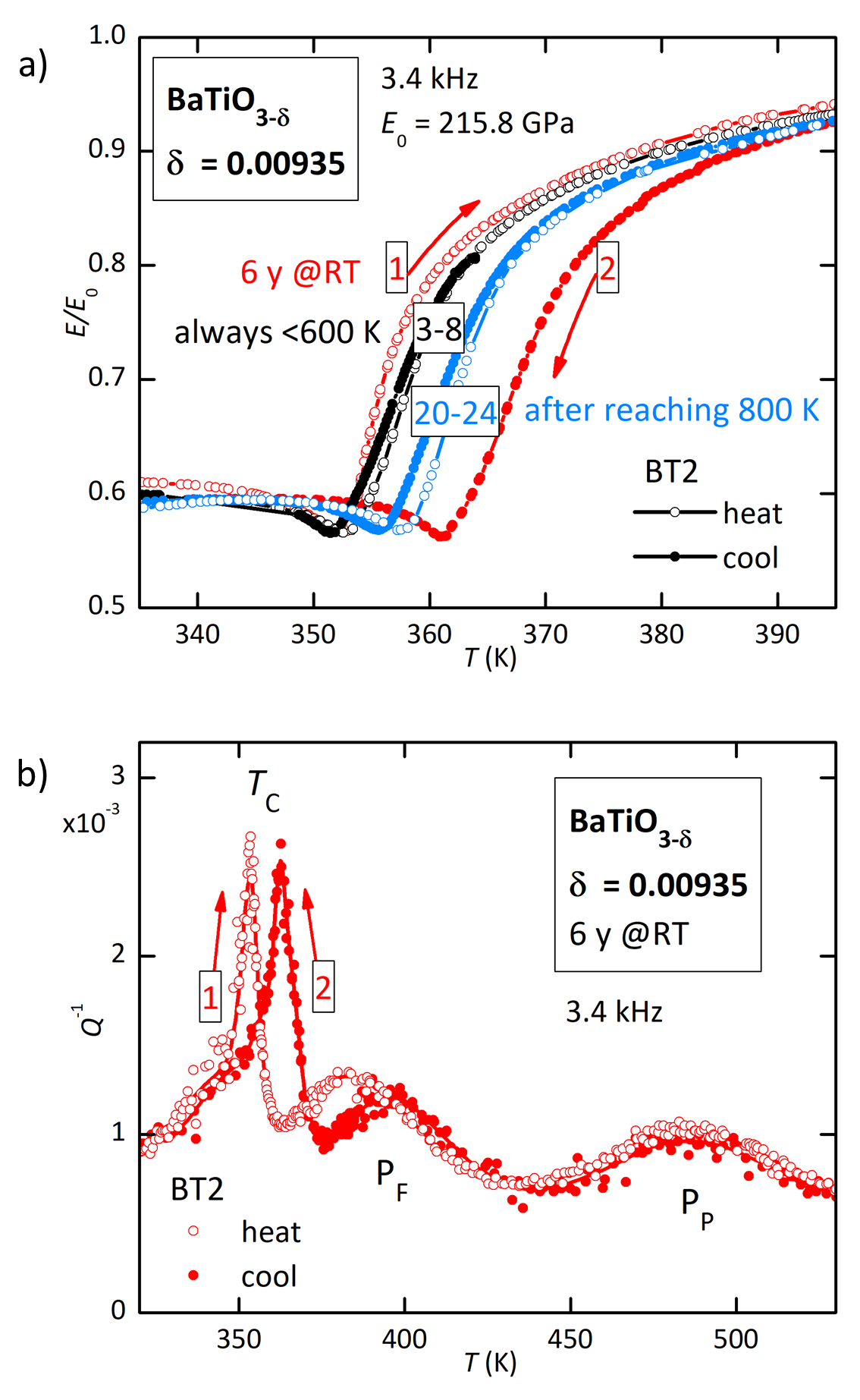}
\caption{a) Normalized Young's modulus of sample BT2 with $\protect\delta =$
0.00935 measured during heating and cooling after 6 years aging at RT (1,2),
subsequent cycles not exceeding 600~K (3-8), subsequent cycles after having
reached 800~K (20-24). b) $Q^{-1}$ measured together with the $E$ of curves
(1,2) in a).}
\label{fig-TcBT6y}
\end{figure}
\end{center}

Figure \ref{fig-TcBT6y}a, however, shows that $T_{\text{C}}$ does not
increase indefinitely and saturate with aging, since the lowest value of $T_{%
\text{C}}$ during heating is measured after aging 6 years, with a huge
inverse thermal hysteresis of 10~K (curves 1 and 2). Then $T_{\text{C}}$
remains stable with normal hysteresis, if cycling does not exceed 600~K
(curves 3-8, only two of which are shown for clarity), but, after reaching
800~K, it increases of 5~K and remains stable there (curves 20-24). Figure %
\ref{fig-TcBT6y}b shows the $Q^{-1}$ curves corresponding to 1 and 2 in Fig. %
\ref{fig-TcBT6y}a, after 6 years aging. Peak P$_{\text{F}}$ during cooling
is shifted to higher temperature, together with $T_{\text{C}}$, and has
smaller intensity.

\subsection{Ba$_{x}$Sr$_{1-x}$TiO$_{3}$}

\begin{center}
\begin{figure}[h]
\includegraphics[width=8.5cm]{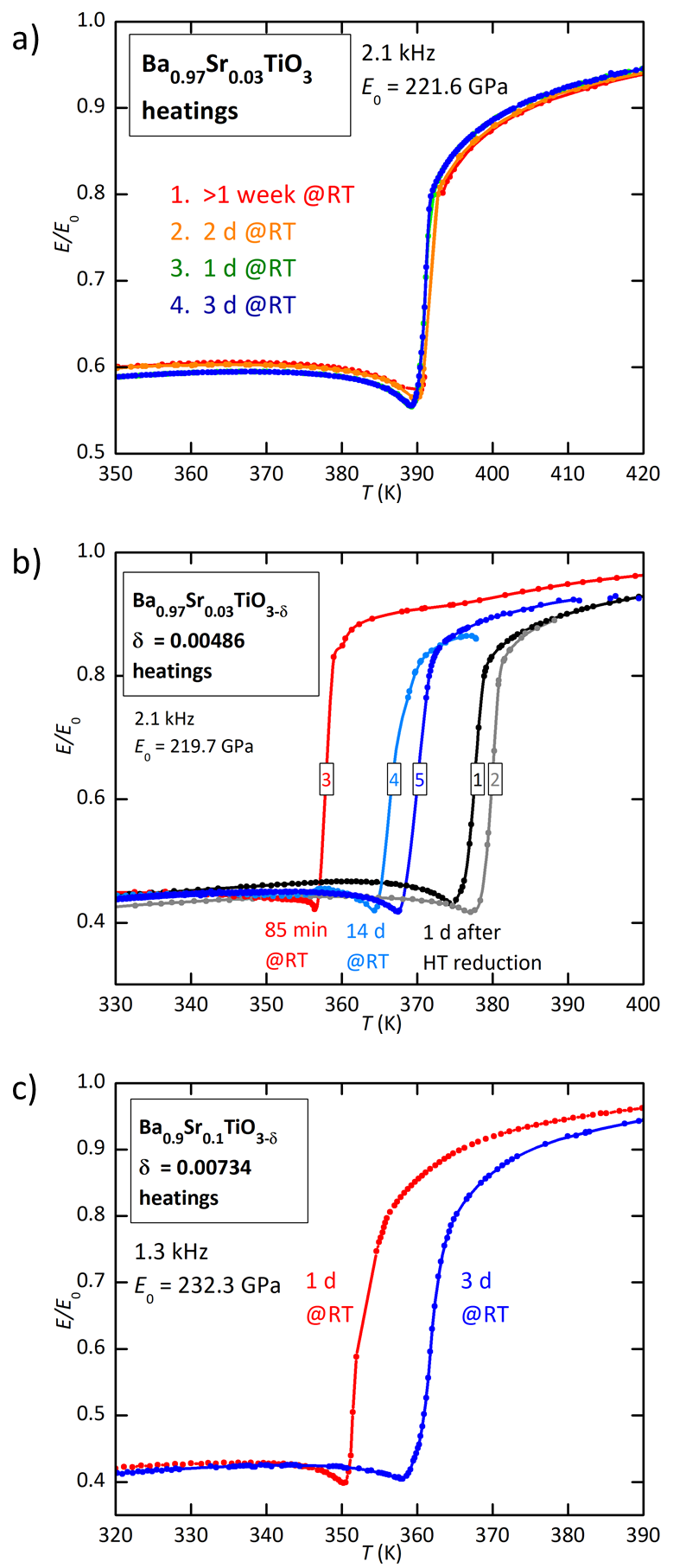}
\caption{Young's modulus of Ba$_{0.9}$Sr$_{0.1}$TiO$_{3-\protect\delta }$
normalized to its maximum value measured during heating with:\ a) $\protect%
\delta =0$ after aging at RT for the times indicated in the figure; b) $%
\protect\delta =$ 0.000486, one day after the reduction treatment (1)
repeated the following day (2), after a permanence of 1.5~h at RT (3), after
14 days at RT (4), immediately cooling past the transition e re-heating (5).}
\label{fig-TcBST}
\end{figure}
\end{center}

Substitution of 3\% Ba with Sr decreases $T_{\text{C}}$ of BaTiO$_{3}$ from
400~K to 390~K, independent of history (Fig. \ref{fig-TcBST}a); yet, upon
introduction of V$_{\text{O}}$ the temperature span of the changes of $T_{%
\text{C}}$ is increased with respect to BaTiO$_{3-\delta }$. Such a span,
only for heating, is $\sim 6$~K in BaTiO$_{3-\delta }$ with $\delta \simeq
0.003$ and 0.009, but is 21~K in Ba$_{0.97}$Sr$_{0.03}$TiO$_{3-\delta }$
with $\delta \simeq $ 0.005 (Fig. \ref{fig-TcBST}b).

Again, there is a trend of higher $T_{\text{C}}$ for longer permanence at RT
in the range of hours and few days, but 14 days aging lowers $T_{\text{C}}$,
reminding the long aging effect in \ref{fig-TcBT6y}a. Few data are available
for $x\left( \text{Sr}\right) =0.1$, but a shift of 7~K in $T_{\text{C}}$
has been observed also there (Fig. \ref{fig-TcBST}c).

\subsection{(Ba$_{0.85}$Ca$_{0.15}$)(Zr$_{0.1}$Ti$_{0.9}$)O$_{3}$}

\begin{center}
\begin{figure}[h]
\includegraphics[width=8.5cm]{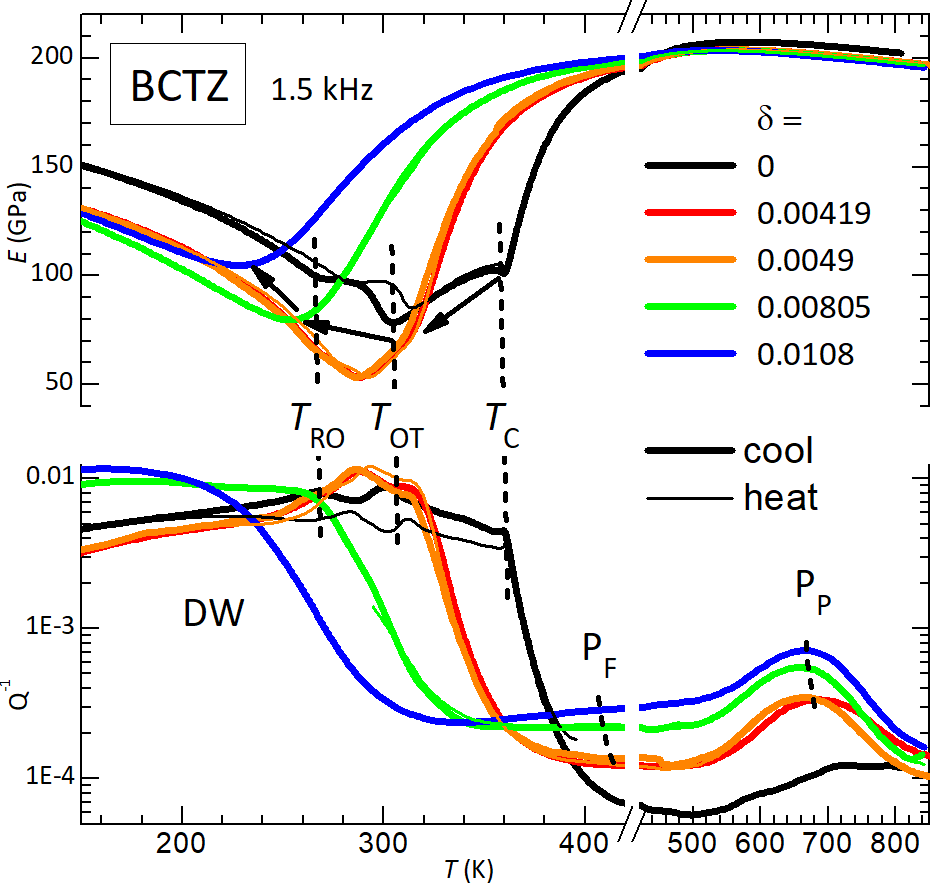}
\caption{Young's modulus $E$ and elastic energy loss coefficient $Q^{-1}$ of
(Ba$_{0.85}$Ca$_{0.15}$)(Zr$_{0.1}$Ti$_{0.9}$)O$_{3-\protect\delta }$ at
various O deficiencies $\protect\delta $. The arrows indicate the shift of $%
T_{\text{C}}$ with increasing $\protect\delta $. The thick lines are
measured during cooling and the thin lines during heating, the difference
being noticeable only at the lowest O deficiencies in the region of the
phase transitions.}
\label{fig-BCTZvsd-ch}
\end{figure}
\end{center}

Figure \ref{fig-BCTZvsd-ch} presents the anelastic spectra of BCTZ at
various O deficiencies, measured during both heating and cooling. The three
phase transitions of BaTiO$_{3}$ from the cubic paraelectric phase to the
tetragonal, orthorhombic and rhombohedral ferroelectric phases are clearly
recognizable in the as prepared oxygenated state. The introduction of V$_{%
\text{O}}$ shifts to lower temperature $T_{\text{C}}$ and $T_{\text{OT}}$,
which are still separated at $\delta \simeq 0.005,$ but beyond that value
the elastic anomalies merge into a broad one.

Also the anelastic relaxation peaks in the paraelectric phase, due to
isolated (P$_{\text{F}}$) and paired (P$_{\text{P}}$) V$_{\text{O}}$ are
recognizable, but there are two major differences with respect to BaTiO$%
_{3-\delta }$ (Fig. \ref{fig-vs-d}):\ the two peaks are shifted to higher
temperature and are much broader.

In order to be more quantitative, we fitted the anelastic spectrum of BCTZ
above $T_{\mathrm{C}}$, following the example of BaTiO$_{3}$ \cite{CTQ21},
with the phenomenological expressions \cite{NB72,Cor93}
\begin{eqnarray}
Q^{-1} &=&\frac{\Delta }{T\cosh ^{2}\left( A/2k_{\text{B}}T\right) }\frac{1}{%
\left( \omega \overline{\tau }\right) ^{\alpha }+\left( \omega \overline{%
\tau }\right) ^{-\beta }}  \label{Eq-peak} \\
\overline{\tau } &=&\tau _{0}\exp \left( W/k_{\text{B}}T\right) /\cosh
\left( A/2k_{\text{B}}T\right)
\end{eqnarray}%
where the usual Debye peak, when $\alpha =\beta =1$, and Arrhenius law for
the relaxation time over a barrier $W$ are modified in order to describe
relaxation between initial and final states differing in energy by $A$, due
to non--equivalence of initial and final state or disorder; this accounts
for intensities of the peaks that do not decrease as $1/T$ with increasing $%
T $. In view of the asymmetric shape of peak P$_{\text{P}}$, two broadening
parameters have been introduced for the low--temperature ($\alpha $) and
high--temperature ($\beta $) sides of the peaks. The high quality anelastic
spectra of SrTiO$_{3-\delta }$ had been fitted with three distinct peaks,
including the intermediate relaxation P$_{\text{I}}$ from the asymmetric
jumps of formation/dissociation of the pairs \cite{Cor07}, and those of BaTiO%
$_{3-\delta }$ with up to five peaks, when distinct peaks appeared at higher
temperatures, attributed to V$_{\text{O}}$ trapped by unwanted defects \cite%
{CTQ21}. In (Ba$_{0.85}$Ca$_{0.15}$)(Zr$_{0.1}$Ti$_{0.9}$)O$_{3} $ the
lattice disorder broadens the relaxation spectrum so much that it is
impossible to distinguish more than two peaks, and therefore the fits were
carried out with two peaks, Eq. (\ref{Eq-peak}), plus a linear background
and an exponential, $A\exp \left( B/T\right) $, to take into account the
rise of dissipation on approaching $T_{\mathrm{C}}$.

\begin{table*}[]
\caption{Fitting parameters of the anelastic spectra of BT (from Ref.
\protect\cite{CTQ21}) and BCTZ in Fig. \protect\ref{fig-fitBCTZ}.}
\label{table:1}%
\begin{tabular}{|r|r|r|r|r|}
\hline
& BT & BT & BCTZ & BCTZ \\ \hline
$\delta $ & 0.0027 & 0.0153 & 0.00419 & 0.0108 \\ \hline
P$_{\text{F}}$: $W_{\text{F}}$ (eV) &  & 0.72 & $0.734$ & $0.804$ \\ \hline
$\tau _{0}$ (s) &  & $2.6\times 10^{-14}$ & $1\times 10^{-13}$ & $2.6\times
10^{-14}$ \\ \hline
$\alpha =\beta $ &  & 0.84 & $0.2$ & $0.2$ \\ \hline
$\tau \left( 293~\text{K}\right) $ &  & 70~ms & 0.4 s & 2 s \\ \hline
P$_{\text{P}}$: $W_{\text{P}}$ (eV) & 0.86 & 0.85 & $1.39$ & $1.39$ \\ \hline
$\tau _{0}$ (s) & $9\times 10^{-14}$ & $4.7\times 10^{-14}$ & $5.3\times
10^{-15}$ & $8.6\times 10^{-15}$ \\ \hline
$\alpha $ & 0.91 & 0.80 & $0.38$ & $0.24$ \\ \hline
$\beta $ & 0.91 & 0.80 & $0.53$ & $1$ \\ \hline
$\tau \left( 293~\text{K}\right) $ & 60~s & 20~s & 130~y & 200~y \\ \hline
\end{tabular}%
\end{table*}

\begin{center}
\begin{figure}[h]
\includegraphics[width=8.5cm]{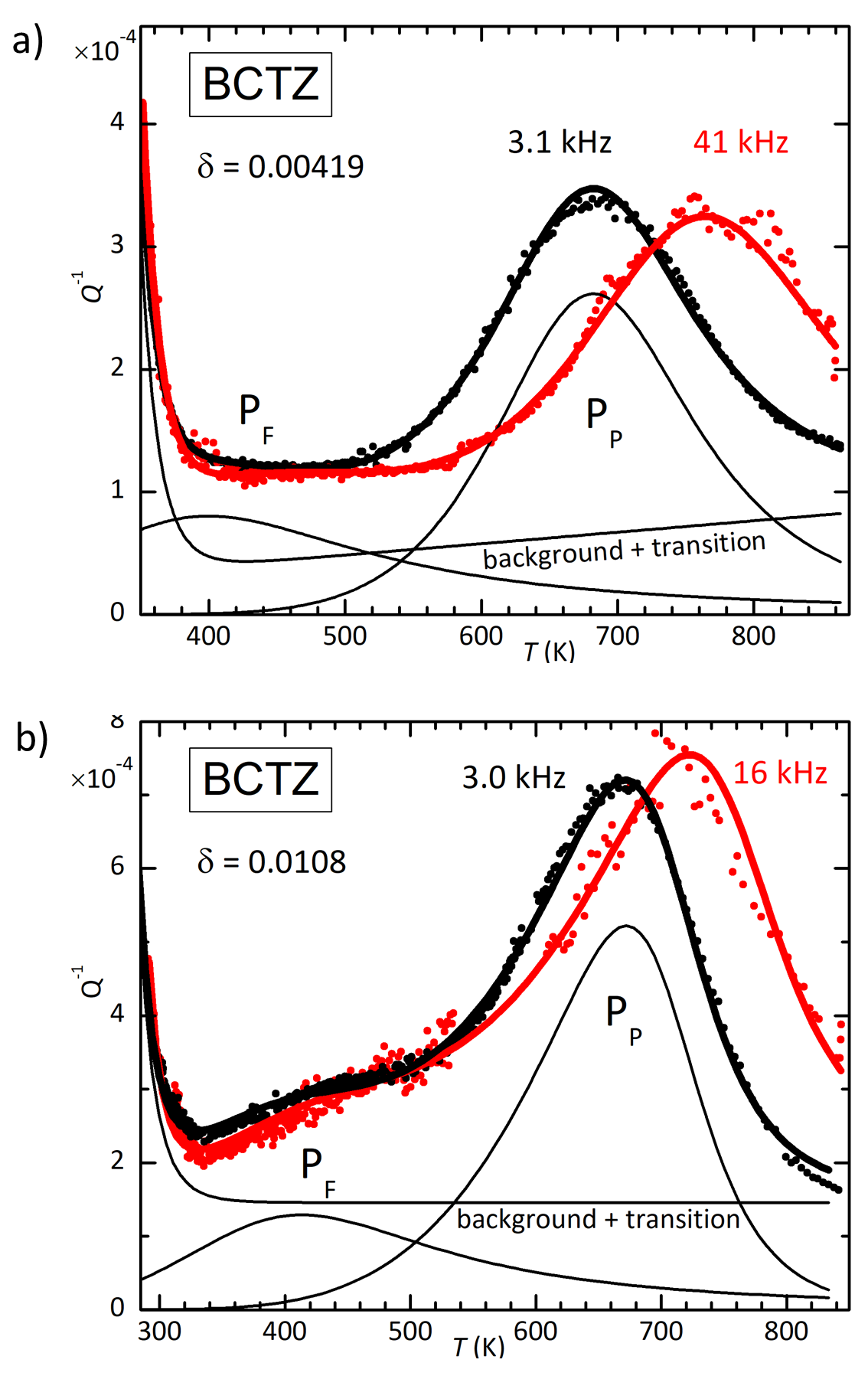}
\caption{The continuous lines are fits of the elastic energy loss
coefficient $Q^{-1}$in the paraelectric phase of O deficient (Ba$_{0.85}$Ca$%
_{0.15}$)(Zr$_{0.1}$Ti$_{0.9}$)O$_{3-\protect\delta }$. a) $\protect\delta =$
0.00419 measured exciting the 1st and 5th modes; b) $\protect\delta =$
0.0108 measured exciting the 1st and 3rd modes.}
\label{fig-fitBCTZ}
\end{figure}
\end{center}

\begin{center}
\begin{figure}[h]
\includegraphics[width=8.5cm]{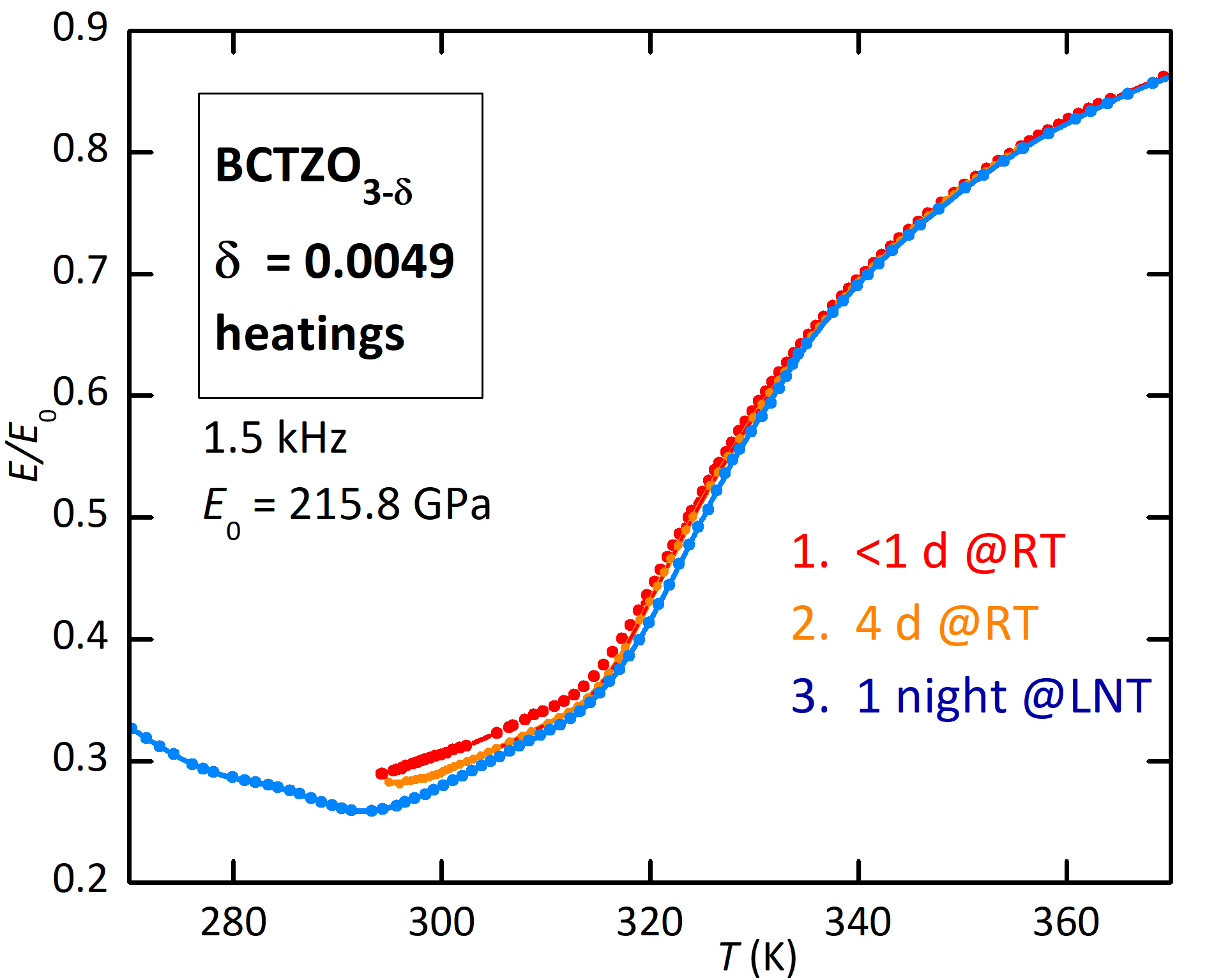}
\caption{Young's modulus $E$ and elastic energy loss coefficient $Q^{-1}$ of
(Ba$_{0.85}$Ca$_{0.15}$)(Zr$_{0.1}$Ti$_{0.9}$)O$_{3-\protect\delta }$ with $%
\protect\delta =$ 0.0049 after various agings in the FE state.}
\label{fig-TcBCTZ}
\end{figure}
\end{center}

Fits to the anelastic spectra of the lowest and highest O deficiencies are
presented in Fig. \ref{fig-fitBCTZ}, while the fitting parameters are
reported in Table I, together with those of two spectra of BaTiO$_{3-\delta
} $ from Ref. \cite{CTQ21}.

Finally, Fig. \ref{fig-TcBCTZ} shows that, while the slight lattice disorder
of 3\% Sr substituting Ba amplifies the dependence of $T_{\text{C}}$ on
aging (Fig. \ref{fig-TcBST} compared with Fig. \ref{fig-TcBT}), the strong
lattice disorder in BCTZ depresses it. The three heating curves differ only
in degree of softening, but have the same $T_{\text{C}}\simeq 315$~K.

\section{Discussion}

The above results show that the V$_{\text{O}}$ not only decrease $T_{\text{C}%
}$, but also make it strongly dependent on history in BT and BST up to 10\%
Sr but not in BCTZ. It should be mentioned that the V$_{\text{O}}$ also
introduce a new phase transition in BaTiO$_{3-\delta }$ slightly above $T_{%
\text{C}}$, which is hardly visible in the anelastic spectra measured at
lower frequency shown here, but is apparent in the modulus measured at
higher frequency. It might be a local effect of stabilization of the
ferroelectric phase around particular clusters of V$_{\text{O}}$, and it
will be the object of a separate study. The additional transition does not
invalidate the fact that there is a major FE transition at a $T_{\text{C}}$
that depends on both $\delta $ and history. We will try to explain these
observations in terms of the most obvious mechanisms that are possible when V%
$_{\text{O}}$ are the only or major defect species: pairing and clustering
of V$_{\text{O}}$ and their association with domain walls.

\subsection{General interpretation of the anelastic spectra}

\subsubsection{Hopping of isolated and paired O vacancies\label{gen}}

\begin{center}
\begin{figure}[h]
\includegraphics[width=8.5cm]{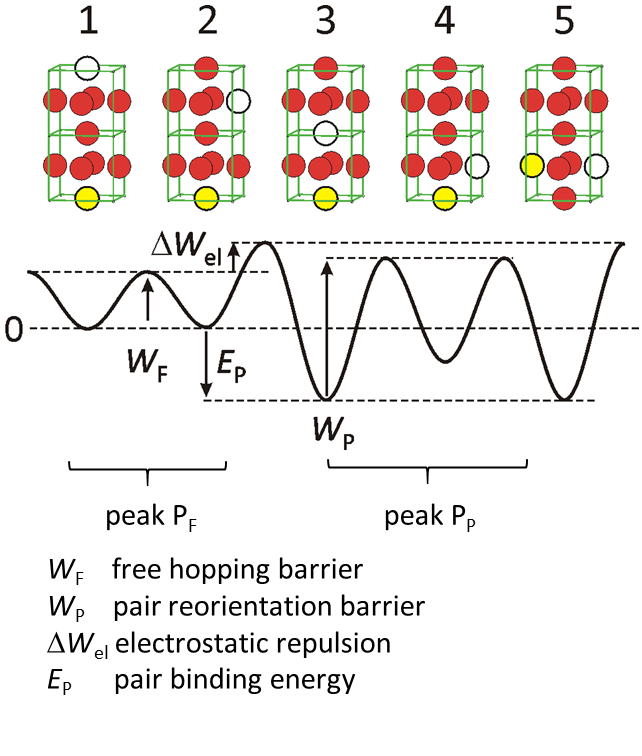}
\caption{Potential profile for passing from configuration 1 to 5, where the
white V$_{\text{O}}$ approaches the yellow one and forms a stable pair (3),
which reorients (4,5).}
\label{fig-pot}
\end{figure}
\end{center}

We refer to previous articles for the detailed analysis and discussion of
the anelastic spectra of SrTiO$_{3-\delta }$ \cite{Cor07} and BaTiO$%
_{3-\delta }$ \cite{CTQ21}, just reminding that the hopping of an V$_{\text{O%
}}$ reorients the associated elastic dipole with major axis parallel to the
nearest neighbor Ti atoms, causing peaks of the form Eq. (\ref{Eq-peak}).
The hopping of isolated V$_{\text{O}}$ causes peak P$_{\mathrm{F}}$, from
which the hopping barrier in SrTiO$_{3}$ is evaluated as $W_{\text{F}}=$
0.60~eV, independent of $\delta $, and in BaTiO$_{3-\delta }$ is $\sim 0.73$%
~eV and decreases with doping. Peak P$_{\mathrm{P}}$ arises from the
reorientation of V$_{\text{O}}$ pairs and has a higher activation energy, $%
W_{\text{P}}=$ $1.0$~eV in SrTiO$_{3-\delta }$, 0.86~eV in BaTiO$_{3-\delta
} $, the latter again decreasing with doping. The lowering of the hopping
barriers with doping is evident from the general shift to lower temperature
of the whole anelastic spectrum of BaTiO$_{3-\delta }$ with increasing $%
\delta $. The activation energy for pair reorientation includes a partial
dissociation of the pair and an increase of the saddle point energy
corresponding to the electrostatic repulsion between the two approaching
charged V$_{\text{O}}$. This effect has been estimated in SrTiO$_{3}$ as a
rise of the saddle point energy for forming a pair by $\Delta W_{\text{el}}=$
0.19~eV with respect to free hopping \cite{Cor07} (see Fig. \ref{fig-pot}).
It causes a slowing of the kinetics for reaching thermal equilibrium between
free and aggregated V$_{\text{O}}$, with respect to what expected from the
high mobility of the isolated V$_{\text{O}}$. A similar effect had been
recognized to play a role with the highly mobile O atoms of the CuO$_{x}$
planes in YBa$_{2}$Cu$_{3}$O$_{6+x}$ \cite{44}.

The actual shape of the potential of the intermediate step 4 for the pair
reorientation is not easy to probe with anelasticity, since the
corresponding peak P$_{\text{I}}$ is depressed by the asymmetry factor in
Eq. (\ref{Eq-peak}), and could be hardly distinguished only in SrTiO$%
_{3-\delta }$ \cite{Cor07}, with a spectrum of nearly pure Debye peaks. Yet,
both the site and saddle point energies should be intermediate between those
of configurations 2 and 3. This means that the complete reorientation
process between 3 and 5, probed by peak P$_{\text{P}}$, has the highest
barrier smaller than that for dissociation from 3 to 2, and \textit{the
dissociation rate is slower than indicated by peak P}$_{\text{P}}$.

The lowering of the hopping barriers at higher O deficiency, observed in
BaTiO$_{3}$ but not in SrTiO$_{3}$, can be explained in terms of a stronger
interaction between the V$_{\text{O}}$ and the nearest neighbor Ti atoms,
when they are off--center \cite{CTQ21}. Indeed, the elastic dipole of an
isolated V$_{\text{O}}$ is three times larger in BaTiO$_{3}$ than in SrTiO$%
_{3}$ and the hopping barrier is 0.73~eV against 0.60~eV. The mobile
electrons doped by the V$_{\text{O}}$ would smear out the eight minima
potential of the Ti atoms, reducing the effect, and making the V$_{\text{O}}$
environment more similar to that of SrTiO$_{3}$, with a shallower potential
\cite{CTQ21}. The doping dependence of the hopping barriers, together with
the facts that the anelastic relaxation peaks of BaTiO$_{3-\delta }$ are
broad and the main peak P$_{\text{F}}$ is visible only at the highest
doping, make challenging an analysis like that carried out for SrTiO$%
_{3-\delta }$, where also the temperature dependence of the populations of
free and aggregated V$_{\text{O}}$ were included and the binding energies
were estimated as $E_{\text{P}}=$ 0.18~eV for pairs and 0.26~eV within
longer chains \cite{Cor07}. Therefore, the fits of the BaTiO$_{3-\delta }$
were carried out with expressions like Eq. (\ref{Eq-peak}), with temperature
independent $\Delta $, and it was assumed that the V$_{\text{O}}$ pair
binding energy in BaTiO$_{3}$ has a value of $\sim 0.2$~eV, similar to SrTiO$%
_{3}$.

The $Q^{-1}\left( T\right) $ curves of BCTZ (Fig. \ref{fig-BCTZvsd-ch}) at
increasing O deficiency are qualitatively similar to those of BaTiO$%
_{3-\delta }$ (Fig. \ref{fig-vs-d}), with the two major peaks P$_{\text{P}}$
and P$_{\text{F}}$ in the paraelectric phase and their shift to lower
temperature with increasing $\delta $, but these peaks are much broader and
have larger activation energies. The broadening, quantified by the
definitely smaller values of $\alpha $ and $\beta $ in Table I, is justified
by the disorder in the cation sizes, and renders difficult the analysis of
the anelastic spectra. Therefore, the parameters of $P_{\text{F}}$ for BCTZ
in Table I are only indicative, and the fact that the activation energies
are not slightly smaller at larger $\delta $, as in BaTiO$_{3-\delta }$,
should not be considered as significant.

Table I contains also the mean characteristic time $\tau $ for the hopping
of an isolated V$_{\text{O}}$ and reorientation of a V$_{\text{O}}$ pair
calculated at room temperature, useful for understanding the mechanisms of
aging, fatigue and deterioration of properties. These times are only
indicative, because neglect the splitting of the energies of the V$_{\text{O}%
}$ with dipole parallel and perpendicular to the tetragonal axis in the FE
phase, as explained later (Fig. \ref{fig-Arrh}). Some points can however be
established: at room temperature the isolated V$_{\text{O}}$ remain quite
mobile, with hopping rates of about $1-10$~s$^{-1}$ in all compositions. The
reorientations of the pairs, and even more their dissociations, are slower,
but still occur over time scales of minutes, except for BCTZ, which stands
out with characteristic times of hundreds of years.\emph{\ }

\begin{center}
\begin{figure}[h]
\includegraphics[width=8.5cm]{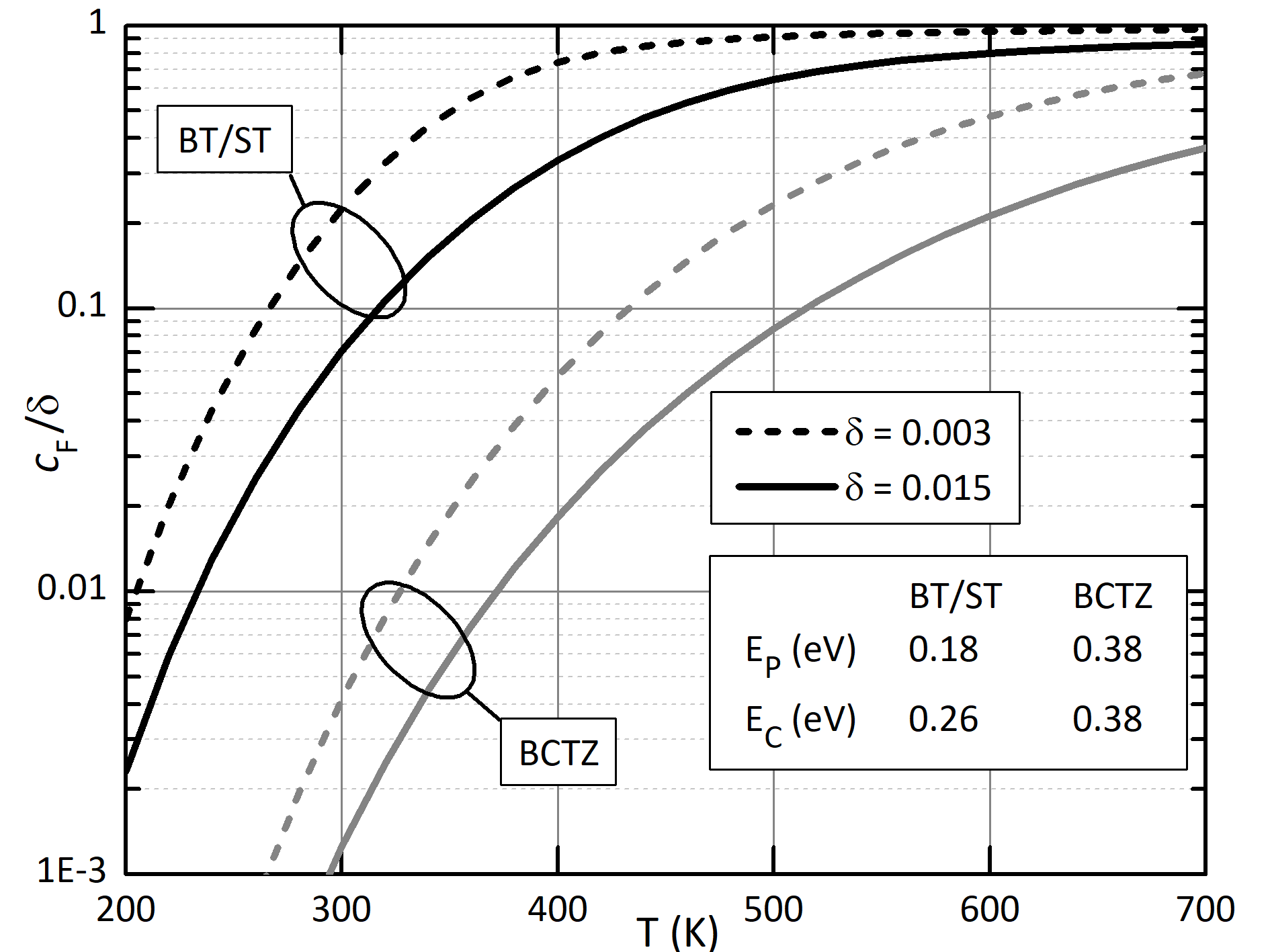}
\caption{Relative fractions of isolated V$_{\text{O}}$ calculated with the
method used for SrTiO$_{3-\protect\delta }$ \protect\cite{Cor07} at two
representative total concentrations $\protect\delta =$ 0.003, 0.015 using a)
the same binding energies of SrTiO$_{3}$, supposed to be valid also for BaTiO%
$_{3}$; b) larger binding energies representative of BCTZ.}
\label{fig-cF}
\end{figure}
\end{center}

Considering the differences in the mobilities~of free and aggregated V$_{%
\text{O}}$, it is important to know their relative fractions as a function
of temperature. As already mentioned, the anelastic spectra of BaTiO$_{3}$
and BCTZ do not allow the binding energies of V$_{\text{O}}$ in pairs and
chains to be estimated, as it was done in SrTiO$_{3}$ \cite{Cor07}. Yet, in
order to have an idea, in Fig. \ref{fig-cF}, the relative fractions of
isolated V$_{\text{O}}$ at two representative total concentrations, $\delta
= $ 0.003 and 0.015, are plotted against $T$ using the same method for
analyzing the anelastic spectra of SrTiO$_{3-\delta }$. In one case, the
same parameters of SrTiO$_{3}$ have been used, which should be
representative of BaTiO$_{3}$, while the other case should represent BCTZ.
Lacking estimates of the pair and chain binding energies for BCTZ, it has
been taken into account that the activation energy for the pair
reorientation in BCTZ, $W_{\text{P}}=1.4$~eV, is 0.4~eV higher than in SrTiO$%
_{3}$ and 0.55~eV higher than in BaTiO$_{3}$, while $W_{\text{F}}$ is higher
by $<0.2$~eV. This suggests that the pair binding energy of BCTZ is $0.2-0.3$%
~eV larger than in SrTiO$_{3}$ and BaTiO$_{3}$, namely $0.38-0.48$~eV. The
curves in Fig. \ref{fig-cF} are plotted using $E_{\text{P}}=$ $E_{\text{C}}=$
0.38~eV, and indicate that, while in BaTiO$_{3-\delta }$ there is a sizeable
fraction of free V$_{\text{O}}$ around $T_{\mathrm{C}}$ and at RT, in BCTZ
practically all V$_{\text{O}}$ are aggregated and therefore static. This
explains why BCTZ does not display any time dependence of $T_{\text{C}}$
(Fig. \ref{fig-TcBCTZ}).

\subsubsection{Pinning of domain walls by O vacancies\label{pin}}

Pinning of DW, at least of the $90^{\circ }$ type, is demonstrated by the
decrease of the dissipation below $T_{\text{C}}$ by $5-10$ times upon
introduction of as little as $\delta \sim $ 0.003 V$_{\text{O}}$ (Fig. \ref%
{fig-vs-d}). This dissipation is due to the relaxational motion of $%
90^{\circ }$ DWs, which enlarge the fraction of the domains with $c$ axis
parallel to the extensional component of the sample vibration strain at the
expenses of those perpendicular to it. There must be a broad distribution of
characteristic times $\tau $ for the DW motion, resulting from the
distribution of domain sizes and configurations, so that, rather than a peak
centred at the temperature where $\omega \tau \sim 1$, the $Q^{-1}$ curve is
flat below $T_{\text{C}}$. It can be concluded that a considerable fraction
of $90^{\circ }$ DWs are pinned already by few tenths of percent of V$_{%
\text{O}}$. Nothing can be said about pinning of the $180^{\circ }$ DWs,
since their motion does not affect strain and does not contribute to the
elastic energy dissipation. Notice, however, that the motion of any type of
DW is also hindered by sparse quasistatic V$_{\text{O}}$ by the fact that,
while the DW sweeps through them, they temporarily transform from apical
with lower energy to equatorial with higher energy (see Sect. \ref{DW}),
effectively acting as a restoring force \cite{CWF16b}.

\subsection{Dependence of $T_{\text{C}}$ on history}

The observation of a time dependence of $T_{\text{C}}$ in ferroelectrics
containing defects is not new. Enhancements of $T_{\text{C}}$ up to 2.5~K
have been observed in single crystals of BaTiO$_{3}$ after aging at RT for
days \cite{SRO05b}, and attributed to the reorientation of Fe$_{\text{Ti}%
}^{\prime }-$V$_{\text{O}}^{\cdot \cdot }$ electric dipoles parallel to the
local spontaneous polarization. The corresponding lowering of the electric
energy would stabilize the ferroelectric domains and enhance $T_{\text{C}}$
\cite{SRO05b}. A related effect is the shift and constriction of the
hysteresis loops, due to the internal field created by the defect dipoles,
preferentially aligned parallel to the local polar direction during aging
\cite{CH77,Ren04}.

Accordingly, the initial slow rise of $T_{\text{C}}$ with aging over hours
and days in the FE state of O deficient BT and BST (Figs. \ref{fig-TcBT}b,c
and \ref{fig-TcBST}b)\ might be partly explained by a stabilization of the
FE domains through the alignment of the orientations of the defect dipoles
with respect to the local polarization, and stabilization of the DW
configurations. In the present situations, the great majority of the V$_{%
\text{O}}$ are introduced by the reduction treatments and are not paired
with acceptor defects to form electric dipoles, but still the interaction of
their elastic, instead of electric, dipoles with the tetragonal strain can
produce the same effect.

\subsection{\protect\bigskip Apical and equatorial V$_{\text{O}}$ in polar
domains}

\begin{center}
\begin{figure}[h]
\includegraphics[width=5.5cm]{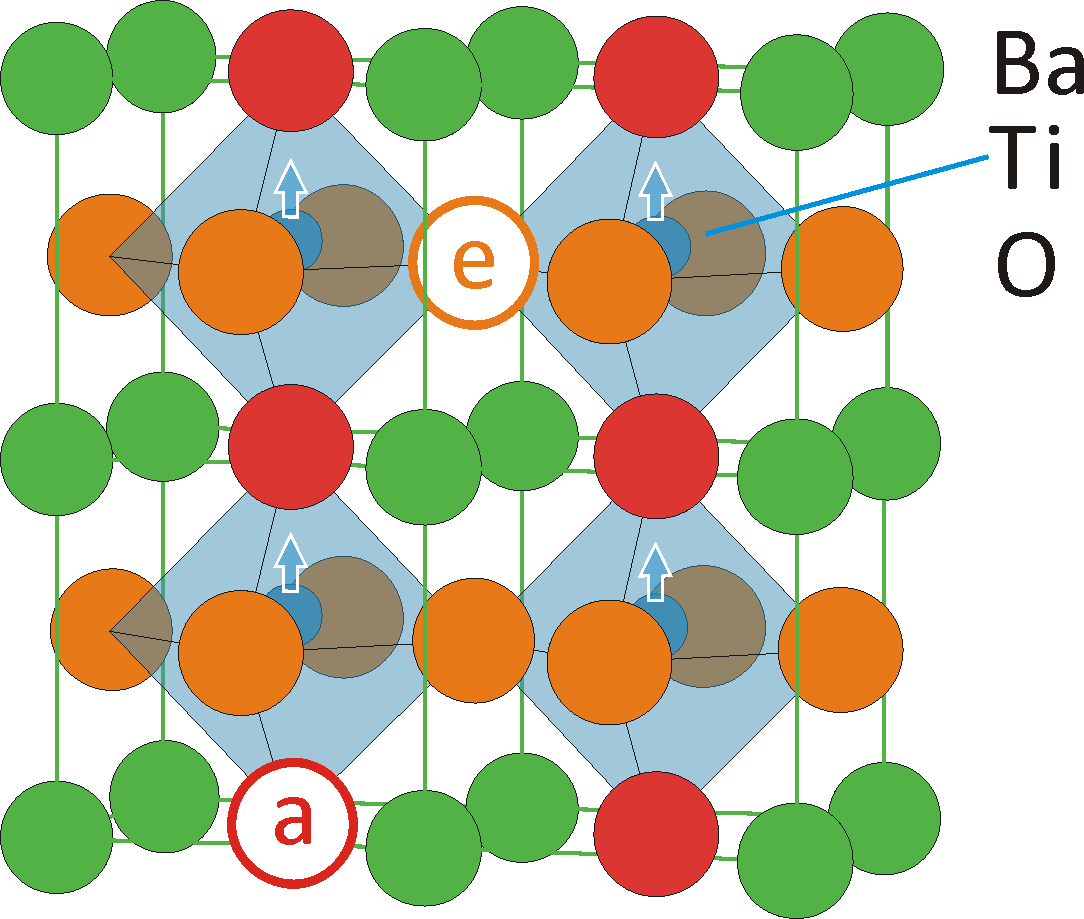}
\caption{Lattice of ferroelectric tetragonal BaTiO$_{3}$ with apical (red)
and equatorial (orange) O$^{2-}$ ions and two respective vacancies. The
arrows are the polar displacements of the Ti atoms.}
\label{fig-aeVO}
\end{figure}
\end{center}

In the FE tetragonal phase of a perovskite the O atoms become of two types:
apical or equatorial with respect to the tetragonal $c$ axis, as shown in
Fig. \ref{fig-aeVO}, and differ in energy by $\Delta E_{\text{ea}}=$ $E_{%
\text{e}}-E_{\text{a}}$, resulting in an interaction with the polar domain.

From a purely elastic point of view it should be $\Delta E_{\text{ea}}>0$,
assuming that the major axis of the V$_{\text{O}}$ strain ellipsoid $\lambda
$ is parallel to the directions of the nearest neighbor Ti$^{4+}$ ions,
pushed away by the positive V$_{\text{O}}$, and in accordance with
calculations \cite{GLE19}. The elastic contribution to $\Delta E_{\mathrm{ea}%
}$ can be estimated as the energy difference of an elastic dipole between
orientations parallel or perpendicular to the tetragonal strain. The
anisotropic component of the elastic dipole of a V$_{\text{O}}$ in cubic
BaTiO$_{3}$, $\Delta \lambda =$ $0.077$, has been deduced from the intensity
of peak P$_{\text{F}}$ \cite{CTQ21} and is very close to the calculated
value of $\Delta \lambda =$ 0.067 \cite{GLE19}. The elastic contribution to
the energy splitting between the two orientations in a tetragonal lattice
can be written as \cite{NB72}
\begin{equation*}
\Delta E_{\text{ea}}^{\text{elas}}=v_{0}\Delta \lambda \left(
c_{33}-c_{13}\right) \varepsilon _{\mathrm{T}}=0.063~\text{eV},
\end{equation*}%
where $v_{0}=$ $64.5\times 10^{-30}~$m$^{3}$ is the cell volume, $c_{11}=211$%
~GPa and $c_{13}=107$~GPa the elastic constants and $\varepsilon _{\mathrm{T}%
}=\frac{c}{a}=0.01$ the tetragonal strain of BaTiO$_{3}$. This value is not
far from $\Delta E_{\text{ea}}=0.1$~eV, calculated for a V$_{\text{O}}$
trapped by a Mn$_{\text{Ti}}$ \cite{NNC15}, and which takes into account
also electronic effects and other differences between the actual FE lattice
and a tetragonally strained PE lattice. Most of the calculations of the
difference between the energies of the apical and equatorial sites for
isolated V$_{\text{O}}$, including electronic effects, have been done for
PbTiO$_{3}$, obtaining quite a broad range of results: $\Delta E_{\text{ea}%
}= $ 0.44~eV \cite{CWF16b}, 0.1~eV \cite{SUW13}, 0.04~eV \cite{YF11},
0.025~eV \cite{XSA16}, $-0.1$~eV \cite{WXX19}, $-0.23$~eV \cite{AF07}. In
the last paper $\Delta E_{\text{ea}} $ is also calculated for BaTiO$_{3}$
and found to be $-0.06$~eV, four times smaller than in PbTiO$_{3}$ \cite%
{AF07}. It is indeed expected that the magnitude of the splitting between
the energies of V$_{\text{O}}$ with elastic dipole parallel and
perpendicular to the $c$ axis is smaller in BaTiO$_{3}$ with $\frac{c}{a}%
=0.01$ than in PbTiO$_{3}$ with $\frac{c}{a}=0.065$.

Summing up, from a purely elastic point of view, the energy of a V$_{\text{O}%
}$ in apical position should be $\Delta E_{\text{ea}}^{\text{elas}}\sim
0.063 $~eV lower than in equatorial position. Taking into account
first--principle calculations, which include the electronic effects, we may
assume that $\Delta E_{\text{ea}}\gtrapprox $ 0.1~eV. This has consequences
with respect to the state and mobility of the V$_{\text{O}}$ at RT, in the
FE state:\ according to the Boltzmann factor $\exp \left( -\Delta E_{\text{ea%
}}/k_{\text{\textrm{B}}}T\right) $, only $\lessapprox 1/50$ of equatorial
sites are occupied by V$_{\text{O}}$, implying almost total alignment of the
elastic dipoles to the tetragonal axis at equilibrium, and the hopping rates
between the two types of positions split by a factor of the same order of
magnitude. These quantities cannot be measured by anelastic relaxation,
because the relaxation spectrum below $T_{\text{C}}$ is dominated by the DW
relaxation.

\begin{center}
\begin{figure}[h]
\includegraphics[width=8.5cm]{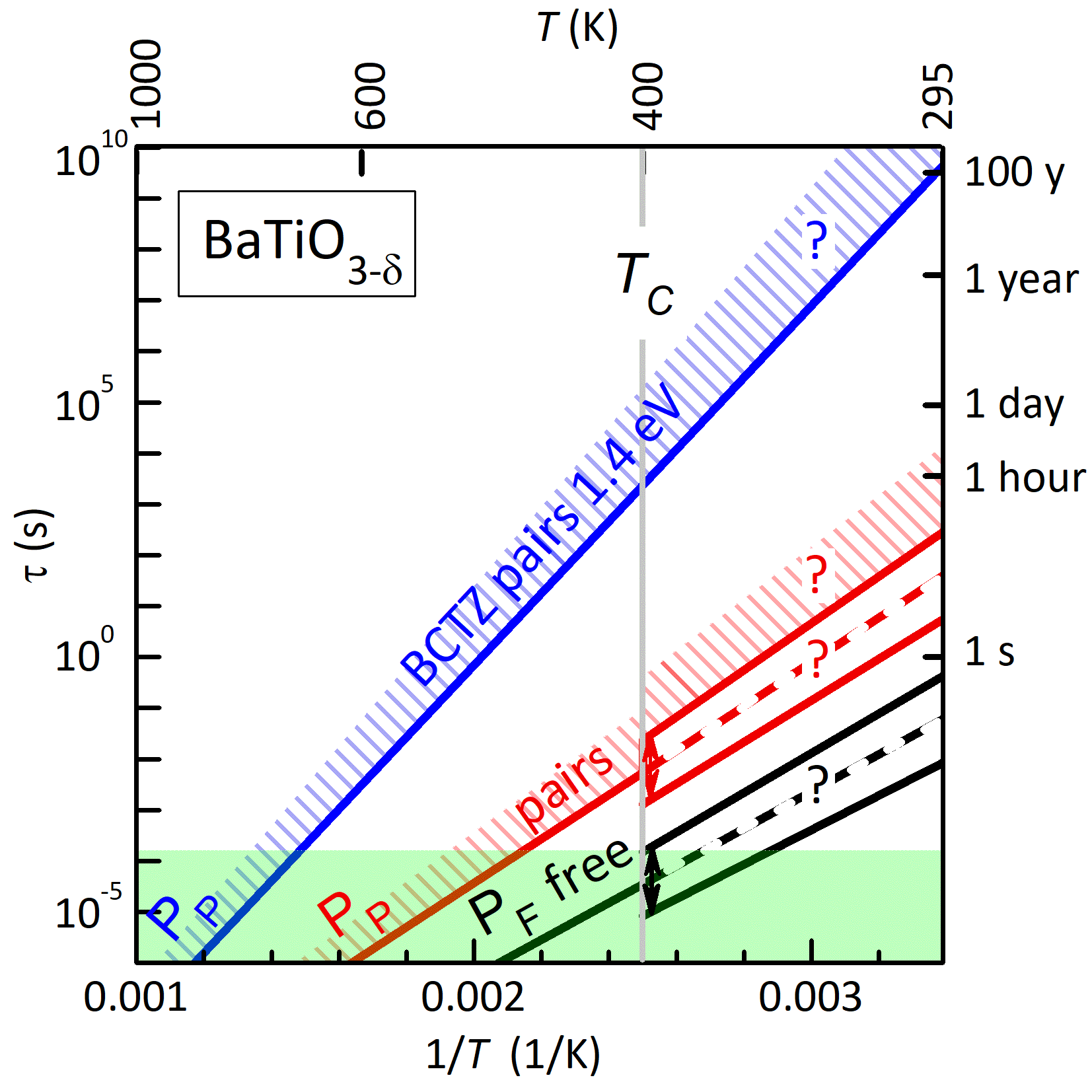}
\caption{Arrhenius plot of the relaxation times of peaks P$_{\text{F}}$ and P%
$_{\text{P}}$ in BT and BCTZ (green region) extrapolated in the FE state. It
is assumed that the splitting of the activation energies in the FE phase is $%
\Delta E_{\text{ea}}=0.1$~eV. The shaded region is the mean pair
dissociation time assuming an activation energy up to 0.1~eV larger than
that for reorientation.}
\label{fig-Arrh}
\end{figure}
\end{center}

The situation is illustrated in Fig. \ref{fig-Arrh}, with an Arrhenius plot
of the relaxation times of peaks P$_{\text{F}}$ and P$_{\text{P}}$ in BT and
BCTZ (green region) extrapolated in the FE state. It is assumed that the
splitting of the activation energies in the FE phase is $\Delta E_{\text{ea}%
}=0.1$~eV. The shaded region is the mean pair dissociation time assuming an
activation energy up to 0.1~eV larger than that for reorientation (see the
discussion of Fig. \ref{fig-pot}).

\subsection{Dependence of electron doping and $T_{\text{C}}$ on the
aggregation of V$_{\text{O}}$\label{doping}}

Having established that the V$_{\text{O}}$ lack electric dipoles, but can
influence the FE state through their elastic dipoles, we still observe that
the magnitude of the changes of $T_{\text{C}}$ with various temperature
protocols in O deficient BT, and especially BST, are much larger than
reported up to now in BT containing dopants or defects. For example, in a
single crystal of BaTiO$_{3}$ with unknown amount of Fe$^{3+}$--V$_{\text{O}%
} $ complexes, $T_{\mathrm{C}}$ increased of up to 2.5~K after aging \cite%
{SRO05b}; in BaTiO$_{3}$ doped with 1\% Mn$^{3+}$ on the Ti$^{4+}$ site, and
presumably $0.5-1\%$ compensating V$_{\text{O}}$, the shift was 1.5~K \cite%
{AYM07}. Evidently, in the presence of uncompensated V$_{\text{O}}$ there
must be another mechanism affecting $T_{\text{C}}$, besides the internal
fields from the oriented defect dipoles. We are going to argue that the
major factor affecting $T_{\mathrm{C}}$ is the electron doping from the
ionized V$_{\text{O}}$, which depends on their aggregation state.

Let us first observe that the major cause of depression of $T_{\text{C}}$ by
doping is not the size effect of the dopants and defects, but the mobile
charges that they induce. This is reasonable, if one considers that the
mobile charges screen the dipolar interactions that induce ferroelectricity,
and is demonstrated by the fact that uncompensated V$_{\text{O}}$ depress $%
T_{\text{C}}$ at a rate of about $-4000$~K/$\delta $ \cite{CTQ21}, but when
they are electrically compensated as Schottky defects \cite{LLK07}, the rate
reduces to $-220$~K/$\delta $ in Ba$_{1-\delta }$TiO$_{3-\delta }$ and $-820$%
~K/$\delta $ in BaTi$_{1-\delta }$O$_{3-2\delta }$, inclusive of the
contribution from the dopants. Accordingly, a neutral defect like Sr in Ba$%
_{1-x}$Sr$_{x}$TiO$_{3}$ reduces $T_{\text{C}}$ by only $-300$~K$/x$ \cite%
{LSS96}. Therefore, considering that the depression of $T_{\text{C}}$ by
charged defects is an order of magnitude larger than that by neutral
defects, it is important to understand what effect the pairing and further
aggregation of V$_{\text{O}}$ has on the electron doping and finally on $T_{%
\text{C}}$.

We will not review the computational studies on the valence state of V$_{%
\text{O}}$ and polarons from ionized V$_{\text{O}}$ in BaTiO$_{3}$, because
there is experimental evidence confirming the existence of V$_{\text{O}}$
pairs and chains, independent of anelastic relaxation \cite{Cor07,CTQ21},
and which also sheds light on the how these clusters affect doping. In fact,
highly reduced films of SrTiO$_{3-\delta }$ have been studied by diffuse
X-ray scattering, finding evidence of linear clusters along the $\left[ 001%
\right] $ direction \cite{ECC17}. Moreover, photoemission spectroscopy
revealed the presence of Ti$^{2+}$, in addition to Ti$^{3+}$ and Ti$^{4+}$
\cite{ECC17}. The $\left[ 001\right] $ linear clusters are the V$_{\text{O}}$
pairs and chains along the O--Ti--O directions, parallel to the polarization
in the FE state, which can also be considered as nuclei of a brownmillerite
phase \cite{CCT16}, and the Ti$^{2+}$ ions are within such pairs and chains
\cite{ECC17}. The Ti$^{3+}$ correspond to the mobile electrons in band or
polaron states, responsible for the nearly metallic conductivity of the
reduced samples, while the electrons at the Ti$^{2+}$ sites are evidently
localized.

The emerging picture is that isolated V$_{\text{O}}$ are doubly charged and
dope two electrons each, but, if two V$_{\text{O}}$ form a pair, doping is
halved, because two electrons are localized on the intermediate Ti$^{2+}$.
Further aggregation of V$_{\text{O}}$ to form longer chains creates new Ti$%
^{2+}$ ions and halves the doping. Therefore, \textit{full
pairing/clustering of the V}$_{\text{O}}$\textit{\ halves the electron
doping, with respect to the case of fully dispersed V}$_{\text{O}}$.

As explained in Sect. \ref{gen}, it is difficult to be quantitative on the
fraction of aggregated V$_{\text{O}}$ during temperature runs and aging for
predicting the consequences on $T_{\text{C}}$, but we may use Fig. \ref%
{fig-cF} as a guide. In addition to the uncertainties on the binding
energies, one should take into account the slow rate for reaching
equilibrium, as mentioned in Sect. \ref{gen}. In fact, the high mobility of
the free V$_{\text{O}}$ ($\sim 10$ jump/s at RT according to Table I)
suggests that the equilibrium fraction of aggregated V$_{\text{O}}$ is
reached almost instantaneously above RT and $T_{\text{C}}$, but the rate for
aggregation is slower than that for free hopping, due to the electrostatic
repulsion, which raises the saddle point for joining another V$_{\text{O}}$
(Fig. \ref{fig-pot}). The rise estimated for SrTiO$_{3}$, $\Delta W_{\text{el%
}}=$ 0.19~eV, causes a slowing of the aggregation the rate by 250 times at
400~K and 2500 times at RT. In addition, if V$_{\text{O}}$ chains have lower
energy than pairs, as in SrTiO$_{3}$, the conversion of sparse pairs into
chains requires the dissociation of some of them, which is slower than the
pair reorientation rate (according to Fig. \ref{fig-Arrh} as slow as $1$
jump/hour at RT). The slow kinetics for reaching the thermal equilibrium
between isolated and aggregated V$_{\text{O}}$ may account for the fact that
it is not sufficient to enter the PE phase in order to reach a stable state,
and both the $Q^{-1}\left( T\right) $ and $E\left( T\right) $ curves,
including $T_{\text{C}}$, are not always reproducible also during cooling.

Since the V$_{\text{O}}$ aggregation decreases doping and enhances $T_{\text{%
C}}$, the slow kinetics for aggregation, may be the major cause of the
increase of $T_{\mathrm{C}}$ with aging in the FE state, and adds to the
known stabilization effects of the FE\ domains. An indication in this sense
comes from the comparison of heating and subsequent cooling after long aging
in Fig. \ref{fig-TcBT6y}b. During heating, peak P$_{\text{F}}$ due to the
hopping of free V$_{\text{O}}$ is more intense, indicating a higher fraction
of free V$_{\text{O}}$ and hence higher doping, and $T_{\mathrm{C}}$ is
indeed lower. In addition, P$_{\text{F}}$ is also shifted to lower
temperature, coherent with higher doping. During cooling the opposite is
observed in peak P$_{\text{F}}$ and $T_{\mathrm{C}}$, coherent with less
free V$_{\text{O}}$ and doping than previously. These phenomena are
amplified by mild lattice disorder, as in BST with 3\% Sr (Fig. \ref%
{fig-TcBST}), presumably due to the distribution of hopping rates with tails
of longer relaxation times.

Explaining the details of the approach to equilibrium during the temperature
runs or short aging times seems an exceedingly difficult task, but we would
now try to explain the puzzling observation that, after an extremely long
aging (6 years in Fig. \ref{fig-TcBT6y}a), $T_{\mathrm{C}}$ does not
actually further increase and saturate, but becomes smaller. In order to do
that, we should consider the competition between the clustering of V$_{\text{%
O}}$ into V$_{\text{O}}$--Ti--V$_{\text{O}}$ pairs and chains along the $%
\left[ 001\right] $ direction and their association to the DWs.

\subsection{Pinning of 90$^{\circ }$ domain walls by isolated V$_{\text{O}}$
and increase of electron doping.\label{DW}}

As noted in Sect. \ref{pin}, the marked depression of $Q^{-1}$ below $T_{%
\text{C}}$ after the introduction of V$_{\text{O}}$, proves that the DW are
pinned by V$_{\text{O}}$. Yet, this does not necessarily imply that the
aggregation state of V$_{\text{O}}$ is affected by DWs. It would be possible
that the direct DW--V$_{\text{O}}$ interaction is weak, the V$_{\text{O}}$
remain within the domains, and pinning is due to the effective restoring
force acting when the DW sweeps through them, and they temporarily transform
from apical with lower energy to equatorial with higher energy \cite{CWF16b}.

In order to establish whether the interaction between V$_{\text{O}}$ and DW
may affect the aggregation state of V$_{\text{O}}$, it is necessary to
compare the respective binding energies and consider in detail which
positions would the V$_{\text{O}}$ occupy with respect to the DW. To this
end, we will review the relevant literature on the nature of the interaction
between V$_{\text{O}}$ and DWs in perovskite ferroelectrics.

In the FE-T phase, the DWs can separate domains whose spontaneous
polarizations are rotated by $180^{\circ }$ and $90^{\circ }$, which in turn
can be neutral head-to-tail DWs or charged head--to--head or tail--to--tail
\cite{OMG13}. Negatively charged tail--to--tail DWs are the most effective
traps for positively charged V$_{\text{O}}^{\cdot \cdot }$, but charged DWs
are energetically unfavorable and therefore rare, unless specific poling
procedures are used \cite{STD12}. This is not the case of our unpoled
samples and we will assume that the DWs are neutral $90^{\circ }$ and $%
180^{\circ }$. On the other hand, it has been proposed that V$_{\text{O}}$
may nucleate tail--to--tail DWs and distribute along them \cite{PKS21}, so
that we cannot exclude that, especially at the highest O deficiencies, also
charged tail--to--tail DW decorated with V$_{\text{O}}$ exist. Even neutral $%
90^{\circ }$ DW, however, have a local electric field, due to the change of
direction of the spontaneous polarization. This field is absent in $%
180^{\circ }$ DW and accumulates positively charged defects and electrons on
either sides of the wall \cite{HSD08}, so that, in general, there is
stronger attraction between V$_{\text{O}}$ and $90^{\circ }$ DWs.

Coming to BaTiO$_{3}$, in most of the experimental studies on pinning of
DWs, the V$_{\text{O}}$ electrically compensate acceptors, like Mn$^{3+}$ or
Fe$^{3+}$ substituting Ti$^{4+}$, intentionally doped or as unwanted
impurities. Since V$_{\text{O}}$ are strongly bound to the acceptors, which
are static, those results are not indicative of DW pinning from mobile V$_{%
\text{O}}$.

First--principle calculations of V$_{\text{O}}$ at both $180^{\circ }$ an $%
90^{\circ }$ DWs have been done mainly for PbTiO$_{3}$. The $180^{\circ }$
DW is found to be extremely sharp, only one unit cell in both directions,
and the energy of a double charged V$_{\text{O}}^{\cdot \cdot }$ at the DW
is 0.13~eV lower than far from it \cite{CDS13}, in substantial agreement
with a previous study \cite{HV03}. In the case of $90^{\circ }$ DW in PbTiO$%
_{3}$, it has been found that both apical and equatorial V$_{\text{O}}$
lower their energy of $\sim 0.2$~eV \cite{CWF16b} or 0.23--0.33~eV \cite%
{XSA16,WXX19} within the DW or in positions nearest neighbor to it. The
first--principle calculations indicate that the sites of lowest energy for
the V$_{\text{O}}$ are on the tail side of $90^{\circ }$ DW, contrary to
what suggested by more macroscopic phase--field calculation schemes \cite%
{HSD08}.

Similar calculations for BaTiO$_{3}$ are not available, except for $%
180^{\circ }$ DW, where it is found that, at variance with PbTiO$_{3}$, V$_{%
\text{O}}$ in BaTiO$_{3}$ have lower energy within the domains than near $%
180^{\circ }$ DW, and concluding that DW pinning by V$_{\text{O}}$ should be
either weak or entirely absent in BaTiO$_{3}$ \cite{SYG21}. This is neither
in contradiction nor in agreement with our data, since we are not sensitive
to the motion of $180^{\circ }$ DW. Moreover, considering that $180^{\circ }$
DWs have a higher formation energy than $90^{\circ }$ DW ($>4$ times in PbTiO%
$_{3}$ \cite{WXX19}), and lack their internal electric fields \cite{HSD08},
the weakness or absence of interaction of V$_{\text{O}}$ with $180^{\circ }$
DW seems irrelevant to our discussion.

Considering that the tetragonal strain and internal electric fields in BaTiO$%
_{3}$ are smaller than in PbTiO$_{3}$, we may take the magnitude of the
binding energies between a V$_{\text{O}}$ and a $90^{\circ }$ DW in PbTiO$%
_{3}$, up to $0.33$~eV, as an upper limit for BaTiO$_{3}$. Therefore, it
appears that the magnitude of the binding energy to the $90^{\circ }$ DW is
competitive with the V$_{\text{O}}$ pair binding energy $E_{\text{P}}\simeq
0.18$~eV in SrTiO$_{3}$ \cite{Cor07}, which we assume is similar, if not
smaller, in BaTiO$_{3}$ \cite{CTQ21}.

\begin{center}
\begin{figure}[h]
\includegraphics[width=8.5cm]{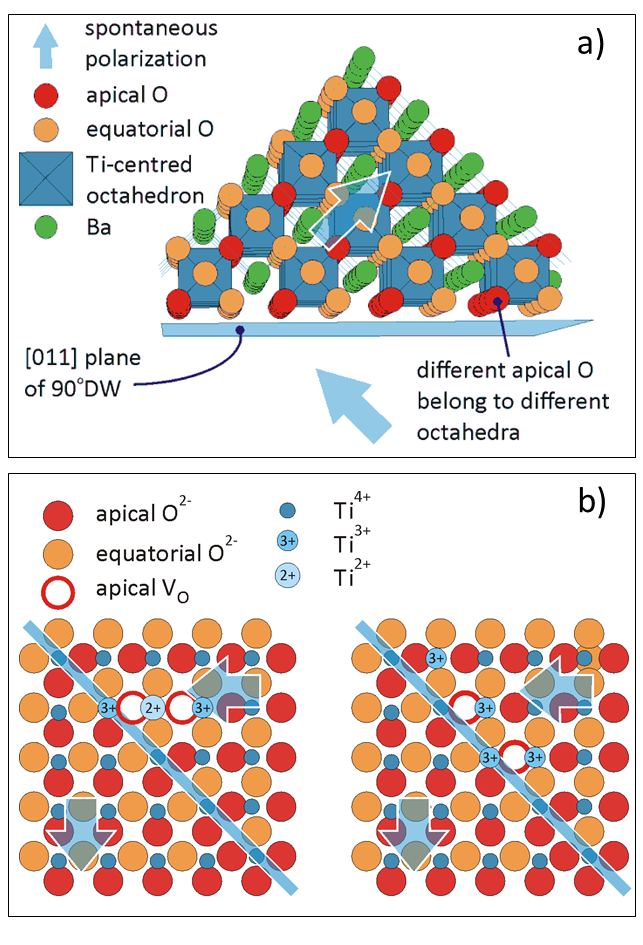}
\caption{a) $90^{\circ }$ DW of BaTiO$_{3}$. All the apical O atoms in
planes parallel to the DW belong to different octahedra. b) TiO$_{2}$ plane
of BaTiO$_{3}$ with $90^{\circ }$ DW (cyan line) and two V$_{\text{O}}$. The
energy of V$_{\text{O}}$ in the equatorial positions is higher than in the
apical positions. On the left, an in-line V$_{\text{O}}$--Ti--V$_{\text{O}}$
pair, where only one V$_{\text{O}}$ has the energy lowered by the DW,
contributes two mobile Ti$^{3+}$ electrons. On the right, the pair is
dissociated in order have both V$_{\text{O}}$ in the lowest energy positions
and contributes four mobile electrons.}
\label{fig-DW}
\end{figure}
\end{center}

It is therefore possible that after a long equilibration time the V$_{\text{O%
}}$ pairs split and migrate to $90^{\circ }$ DWs, where the most favorable
configuration is no more that of stable V$_{\text{O}}$--Ti--V$_{\text{O}}$
pairs and chains. This is evident in Fig. \ref{fig-DW}a, showing a domain
ending at the bottom in a $90^{\circ }$ DW, with the TiO$_{6}$ octahedra put
in evidence. The O atoms are distinguished into apical and equatorial. We
have seen that the Boltzmann factor $\exp \left( -\Delta E_{\text{ea}}/k_{%
\text{\textrm{B}}}T\right) $ at RT makes an apical V$_{\text{O}}\ $tens of
time more favorable than an equatorial one, both far and near a DW.
Therefore, the V$_{\text{O}}$ occupy only the apical positions, both when
they are isolated and aggregated, and the stable V$_{\text{O}}$ pairs and
chains in the FE state are along these $\left[ 001\right] $ O--Ti--O rows,
but it is clear that no such pair can have both V$_{\text{O}}$ at the DW
plane. In fact, if a pair is along the $c$ direction, within the horizontal
plane of Fig. \ref{fig-DW}a, then only one of the two V$_{\text{O}}$ is at
the DW, and the DFT calculations show very sharp differences between the
optimal position at the DW and away from it \cite{XSA16,WXX19}. Neither can
any pair of sites in the plane of the DW form a stable pair of V$_{\text{O}}$
at opposite sides of a same Ti atom, since different apical O sites belong
to different octahedra.

Figure \ref{fig-DW}b shows what happens when a stable pair of V$_{\text{O}}$%
, sharing a Ti$^{2+}$ ion, dissociates in order to have both V$_{\text{O}}$
at the DW: the number of mobile Ti$^{3+}$ electrons doubles from 2 to 4, as
explained in Sect. \ref{doping}.

In the light of the above considerations, we propose the following
explanation for the fact that $T_{\text{C}}$ increases with aging on the
time scale of days but then decreases after years (Fig. \ref{fig-TcBT6y}a).
When cooling from the PE state through $T_{\text{C}}$, about half of the V$_{%
\text{O}}$ are free (Fig. \ref{fig-cF}) and, in addition to continue
aggregating, they can quickly pin the forming DWs, as demonstrated by the
fact that the elastic energy loss of the O deficient samples is depressed
immediately below $T_{\text{C}}$, with respect to the case $\delta =0$. The
first stage of aging at RT would see a clamping of the DWs mainly by free V$%
_{\text{O}}$, which does not change doping, and an increase of the
aggregated fraction far from the DWs, according to Fig. \ref{fig-cF}, which
decreases doping and raises $T_{\text{C}}$. Yet, if the V$_{\text{O}}$ are
slightly more stable at the DWs than in pairs and chains, which seems to be
the case, on the long time, part of the already aggregated V$_{\text{O}}$
will dissociate and migrate to the DWs, accounting for the final decrease of
$T_{\text{C}}$. The relaxation times extrapolated at RT in Fig. \ref%
{fig-Arrh}, together with the equilibrium fractions of isolated and
aggregated V$_{\text{O}}$ in Fig. \ref{fig-cF}, justify the observation of
the reach phenomenology found in the aging dependence of $T_{\text{C}}$ in
BT and BST.

\subsection{Absence of aging in BCTZ}

Figure \ref{fig-TcBCTZ} demonstrates that, contrary to BT and BST, aging has
no effect on $T_{\mathrm{C}}$ in BCTZ. This is again explained by the
combination of Figs. \ref{fig-cF} and \ref{fig-Arrh}: at RT practically all V%
$_{\text{O}}$ are aggregated and their mean dissociation rate is of the
order of hundreds of years. This may also provide a justification to the
better resistance to bipolar fatigue \cite{FKR18} and absence of pinched P-E
loops starting from the unpoled state found in BCTZ \cite{ZGE15}.

\section{Conclusions}

The Young's modulus and elastic energy loss, $Q^{-1}$, of BaTiO$_{3-\delta }$%
, Ba$_{x}$Sr$_{1-x}$TiO$_{3-\delta }$ and (Ba$_{0.85}$Ca$_{0.15}$)(Zr$_{0.1}$%
Ti$_{0.9}$)O$_{3-\delta }$ have been repeatedly measured during heating and
cooling runs at various O deficiencies and after different aging times. The $%
Q^{-1}\left( T\right) $ curves in the paraelectric phase contain peaks due
to the isolated and paired V$_{\text{O}}$, from which it is possible to
measure the respective hopping/reorientation rates and activation energies,
while in the ferroelectric phase they monitor the degree of domain wall
pinning by V$_{\text{O}}$.

The V$_{\text{O}}$ are introduced by reducing treatments and are free to
diffuse and aggregate into pairs, while in most applications they are
compensating acceptor dopants, with which form complexes stable at room
temperature. A major consequence is that the dependence of $T_{\text{C}}$
and its thermal hysteresis on aging and history are greatly enhanced: in BT
(BST) we found that $T_{\text{C}}$ measured during heating may vary of 6~K
(21~K) at the same value of $\delta $, while no anomaly is found in BCTZ.
The main conclusions are that $T_{\text{C}}$ is depressed mainly by the
mobile electrons doped by V$_{\text{O}}$. Each isolated V$_{\text{O}}$ dopes
two mobile electrons as Ti$^{3+}$ ions, but, when it forms a stable linear V$%
_{\text{O}}$--Ti$^{2+}$--V$_{\text{O}}$ pair, the two electrons of the Ti$%
^{2+}$ ion are immobile and do not contribute to doping. Therefore, $T_{%
\text{C}}$ is determined not only by the O deficiency $\delta $ but also by
the fraction of V$_{\text{O}}$ that are aggregated. In this manner it is
possible to explain the rise of $T_{\text{C}}$ during initial aging, much
larger than in acceptor doped BaTiO$_{3}$, where the V$_{\text{O}}$ remain
bound to the acceptor and do not vary doping.

While a mild lattice disorder (up to 10\% Sr substituting Ba) increases the
effects, the strong disorder of BCTZ completely freezes them. In fact, BCTZ
has a much larger mean activation energy for the pair reorientation (1.4~eV
compared to $\leq 0.86$~eV in BT) and hence for the pair dissociation. At
room temperature practically all V$_{\text{O}}$ are paired and static over a
time scale of hundreds of years, explaining the superior resistance of BCTZ
to fatigue \cite{FKR18}.

In order to explain the fact that during long aging (here 6 years) $T_{\text{%
C}}$ does not continue increasing and saturate, but actually decreases, the
interaction between V$_{\text{O}}$ and $90^{\circ }$ DW has been considered
in detail. It is shown that the stable linear V$_{\text{O}}$--Ti--V$_{\text{O%
}}$ pairs along the polarization direction cannot optimize their position
with respect to the DW plane. Therefore, over a long time scale, the V$_{%
\text{O}}$ initially in stable pairs, dissociate in order to decorate the $%
90^{\circ }$ DW, increasing doping and lowering $T_{\text{C}}$.

\section*{Acknowledgments}

This work has been partially funded by the PRIN SELWA Project, Next
Generation EU n. 20229PNWM7 and the S\~{a}o Paulo Research Foundation
FAPESP, Brazil (Grant No. 2022/08030-5). FC and FC  acknowledge the precious
technical assistance of M.P. Latino (CNR-ISM).

\bibliographystyle{apsrev4-1}
\bibliography{refs}

\begin{thebibliography}{46}%
\makeatletter
\providecommand \@ifxundefined [1]{%
 \@ifx{#1\undefined}
}%
\providecommand \@ifnum [1]{%
 \ifnum #1\expandafter \@firstoftwo
 \else \expandafter \@secondoftwo
 \fi
}%
\providecommand \@ifx [1]{%
 \ifx #1\expandafter \@firstoftwo
 \else \expandafter \@secondoftwo
 \fi
}%
\providecommand \natexlab [1]{#1}%
\providecommand \enquote  [1]{``#1''}%
\providecommand \bibnamefont  [1]{#1}%
\providecommand \bibfnamefont [1]{#1}%
\providecommand \citenamefont [1]{#1}%
\providecommand \href@noop [0]{\@secondoftwo}%
\providecommand \href [0]{\begingroup \@sanitize@url \@href}%
\providecommand \@href[1]{\@@startlink{#1}\@@href}%
\providecommand \@@href[1]{\endgroup#1\@@endlink}%
\providecommand \@sanitize@url [0]{\catcode `\\12\catcode `\$12\catcode
  `\&12\catcode `\#12\catcode `\^12\catcode `\_12\catcode `\%12\relax}%
\providecommand \@@startlink[1]{}%
\providecommand \@@endlink[0]{}%
\providecommand \url  [0]{\begingroup\@sanitize@url \@url }%
\providecommand \@url [1]{\endgroup\@href {#1}{\urlprefix }}%
\providecommand \urlprefix  [0]{URL }%
\providecommand \Eprint [0]{\href }%
\providecommand \doibase [0]{http://dx.doi.org/}%
\providecommand \selectlanguage [0]{\@gobble}%
\providecommand \bibinfo  [0]{\@secondoftwo}%
\providecommand \bibfield  [0]{\@secondoftwo}%
\providecommand \translation [1]{[#1]}%
\providecommand \BibitemOpen [0]{}%
\providecommand \bibitemStop [0]{}%
\providecommand \bibitemNoStop [0]{.\EOS\space}%
\providecommand \EOS [0]{\spacefactor3000\relax}%
\providecommand \BibitemShut  [1]{\csname bibitem#1\endcsname}%
\let\auto@bib@innerbib\@empty
\bibitem [{\citenamefont {Genenko}\ \emph {et~al.}(2015)\citenamefont
  {Genenko}, \citenamefont {Glaum}, \citenamefont {Hoffmann},\ and\
  \citenamefont {Albe}}]{GGH15}%
  \BibitemOpen
  \bibfield  {author} {\bibinfo {author} {\bibfnamefont {Y.~A.}\ \bibnamefont
  {Genenko}}, \bibinfo {author} {\bibfnamefont {J.}~\bibnamefont {Glaum}},
  \bibinfo {author} {\bibfnamefont {M.~J.}\ \bibnamefont {Hoffmann}}, \ and\
  \bibinfo {author} {\bibfnamefont {K.}~\bibnamefont {Albe}},\ }\bibfield
  {title} {\emph {\enquote {\bibinfo {title} {{Mechanisms of aging and fatigue
  in ferroelectrics}},}\ }}\href {\doibase 10.1016/j.mseb.2014.10.003}
  {\bibfield  {journal} {\bibinfo  {journal} {Mater. Sci. Engin. B}\ }\textbf
  {\bibinfo {volume} {192}},\ \bibinfo {pages} {52} (\bibinfo {year}
  {2015})}\BibitemShut {NoStop}%
\bibitem [{\citenamefont {Tyunina}(2020)}]{Tyu20}%
  \BibitemOpen
  \bibfield  {author} {\bibinfo {author} {\bibfnamefont {M.}~\bibnamefont
  {Tyunina}},\ }\bibfield  {title} {\emph {\enquote {\bibinfo {title} {{Oxygen
  Vacancies in Perovskite Oxide Piezoelectrics}},}\ }}\href {\doibase
  10.3390/ma13245596} {\bibfield  {journal} {\bibinfo  {journal} {Materials}\
  }\textbf {\bibinfo {volume} {13}},\ \bibinfo {pages} {5596} (\bibinfo {year}
  {2020})}\BibitemShut {NoStop}%
\bibitem [{\citenamefont {Chun}\ \emph {et~al.}(2024)\citenamefont {Chun},
  \citenamefont {Heo}, \citenamefont {Lee}, \citenamefont {Ye},\ and\
  \citenamefont {Yoon}}]{CHL24}%
  \BibitemOpen
  \bibfield  {author} {\bibinfo {author} {\bibfnamefont {J.}~\bibnamefont
  {Chun}}, \bibinfo {author} {\bibfnamefont {J.}~\bibnamefont {Heo}}, \bibinfo
  {author} {\bibfnamefont {K.}~\bibnamefont {Lee}}, \bibinfo {author}
  {\bibfnamefont {B.~U.}\ \bibnamefont {Ye}}, \ and\ \bibinfo {author}
  {\bibfnamefont {B.~S. K. S.-H.}\ \bibnamefont {Yoon}},\ }\bibfield  {title}
  {\emph {\enquote {\bibinfo {title} {{Thermal activation energy on electrical
  degradation process in BaTiO$_3$ based multilayer ceramic capacitors for
  lifetime reliability}},}\ }}\href {\doibase 10.1038/s41598-024-51254-w}
  {\bibfield  {journal} {\bibinfo  {journal} {Sci. Rep.}\ }\textbf {\bibinfo
  {volume} {14}},\ \bibinfo {pages} {616} (\bibinfo {year} {2024})}\BibitemShut
  {NoStop}%
\bibitem [{\citenamefont {Opitz}\ \emph {et~al.}(2003)\citenamefont {Opitz},
  \citenamefont {Albertsen}, \citenamefont {Beeson}, \citenamefont {Hennings},
  \citenamefont {Routbort},\ and\ \citenamefont {Randall}}]{OAB03}%
  \BibitemOpen
  \bibfield  {author} {\bibinfo {author} {\bibfnamefont {M.~R.}\ \bibnamefont
  {Opitz}}, \bibinfo {author} {\bibfnamefont {K.}~\bibnamefont {Albertsen}},
  \bibinfo {author} {\bibfnamefont {J.~J.}\ \bibnamefont {Beeson}}, \bibinfo
  {author} {\bibfnamefont {D.~F.}\ \bibnamefont {Hennings}}, \bibinfo {author}
  {\bibfnamefont {J.~L.}\ \bibnamefont {Routbort}}, \ and\ \bibinfo {author}
  {\bibfnamefont {C.~A.}\ \bibnamefont {Randall}},\ }\bibfield  {title} {\emph
  {\enquote {\bibinfo {title} {{Kinetic Process of Reoxidation of Base Metal
  Technology BaTiO$_3$-Based Multilayer Capacitors}},}\ }}\href {\doibase
  10.1111/j.1151-2916.2003.tb03576.x} {\bibfield  {journal} {\bibinfo
  {journal} {J. Am. Ceram. Soc.}\ }\textbf {\bibinfo {volume} {86}},\ \bibinfo
  {pages} {1879} (\bibinfo {year} {2003})}\BibitemShut {NoStop}%
\bibitem [{\citenamefont {Zhang}\ \emph {et~al.}(2024)\citenamefont {Zhang},
  \citenamefont {Tan}, \citenamefont {Wang}, \citenamefont {Huang},
  \citenamefont {Lin}, \citenamefont {Huang}, \citenamefont {Yang},
  \citenamefont {Fu}, \citenamefont {Cao}, \citenamefont {Zhang}, \citenamefont
  {Yu},\ and\ \citenamefont {Sun}}]{ZTW24}%
  \BibitemOpen
  \bibfield  {author} {\bibinfo {author} {\bibfnamefont {F.}~\bibnamefont
  {Zhang}}, \bibinfo {author} {\bibfnamefont {J.}~\bibnamefont {Tan}}, \bibinfo
  {author} {\bibfnamefont {P.}~\bibnamefont {Wang}}, \bibinfo {author}
  {\bibfnamefont {R.}~\bibnamefont {Huang}}, \bibinfo {author} {\bibfnamefont
  {H.-T.}\ \bibnamefont {Lin}}, \bibinfo {author} {\bibfnamefont
  {X.}~\bibnamefont {Huang}}, \bibinfo {author} {\bibfnamefont
  {J.}~\bibnamefont {Yang}}, \bibinfo {author} {\bibfnamefont {Z.}~\bibnamefont
  {Fu}}, \bibinfo {author} {\bibfnamefont {X.}~\bibnamefont {Cao}}, \bibinfo
  {author} {\bibfnamefont {L.}~\bibnamefont {Zhang}}, \bibinfo {author}
  {\bibfnamefont {S.}~\bibnamefont {Yu}}, \ and\ \bibinfo {author}
  {\bibfnamefont {R.}~\bibnamefont {Sun}},\ }\bibfield  {title} {\emph
  {\enquote {\bibinfo {title} {{Defect dipole engineering enhanced the
  dielectric performance and reliability of Mn-doped BaTiO$_3$-based multilayer
  ceramic capacitor}},}\ }}\href {\doibase 10.1016/j.ceramint.2024.07.189}
  {\bibfield  {journal} {\bibinfo  {journal} {Ceram. Int.}\ }\textbf {\bibinfo
  {volume} {50}},\ \bibinfo {pages} {38263} (\bibinfo {year}
  {2024})}\BibitemShut {NoStop}%
\bibitem [{\citenamefont {Huangfu}\ \emph {et~al.}(2024)\citenamefont
  {Huangfu}, \citenamefont {Wang}, \citenamefont {Zhang}, \citenamefont {Chen},
  \citenamefont {Liu},\ and\ \citenamefont {Guo}}]{HWZ24}%
  \BibitemOpen
  \bibfield  {author} {\bibinfo {author} {\bibfnamefont {G.}~\bibnamefont
  {Huangfu}}, \bibinfo {author} {\bibfnamefont {J.}~\bibnamefont {Wang}},
  \bibinfo {author} {\bibfnamefont {H.}~\bibnamefont {Zhang}}, \bibinfo
  {author} {\bibfnamefont {J.}~\bibnamefont {Chen}}, \bibinfo {author}
  {\bibfnamefont {Z.}~\bibnamefont {Liu}}, \ and\ \bibinfo {author}
  {\bibfnamefont {Y.}~\bibnamefont {Guo}},\ }\bibfield  {title} {\emph
  {\enquote {\bibinfo {title} {{Deciphering the Effect of Defect Dipoles on the
  Polarization and Electrostrain Behavior in Perovskite Ferroelectrics}},}\
  }}\href {\doibase 10.1021/acs.nanolett.4c03042} {\bibfield  {journal}
  {\bibinfo  {journal} {Nano Lett.}\ }\textbf {\bibinfo {volume} {24}},\
  \bibinfo {pages} {12148} (\bibinfo {year} {2024})}\BibitemShut {NoStop}%
\bibitem [{\citenamefont {Zheng}\ \emph {et~al.}(2022)\citenamefont {Zheng},
  \citenamefont {Sun}, \citenamefont {Qi}, \citenamefont {Yang}, \citenamefont
  {Zhang},\ and\ \citenamefont {Cao}}]{ZSQ22}%
  \BibitemOpen
  \bibfield  {author} {\bibinfo {author} {\bibfnamefont {H.}~\bibnamefont
  {Zheng}}, \bibinfo {author} {\bibfnamefont {E.}~\bibnamefont {Sun}}, \bibinfo
  {author} {\bibfnamefont {X.}~\bibnamefont {Qi}}, \bibinfo {author}
  {\bibfnamefont {B.}~\bibnamefont {Yang}}, \bibinfo {author} {\bibfnamefont
  {R.}~\bibnamefont {Zhang}}, \ and\ \bibinfo {author} {\bibfnamefont
  {W.}~\bibnamefont {Cao}},\ }\bibfield  {title} {\emph {\enquote {\bibinfo
  {title} {{Temperature and frequency dependent defect dipole kinematics in
  "hard" piezoelectric ceramics}},}\ }}\href {\doibase
  10.1016/j.sna.2022.113712} {\bibfield  {journal} {\bibinfo  {journal} {Sens.
  Actuators A: Phys.}\ }\textbf {\bibinfo {volume} {344}},\ \bibinfo {pages}
  {113712} (\bibinfo {year} {2022})}\BibitemShut {NoStop}%
\bibitem [{\citenamefont {Souza}(2015)}]{DeS15}%
  \BibitemOpen
  \bibfield  {author} {\bibinfo {author} {\bibfnamefont {R.~A.~D.}\
  \bibnamefont {Souza}},\ }\bibfield  {title} {\emph {\enquote {\bibinfo
  {title} {{Oxygen Diffusion in SrTiO$_3$ and Related Perovskite Oxides}},}\
  }}\href {\doibase 10.1002/adfm.201500827} {\bibfield  {journal} {\bibinfo
  {journal} {Adv. Func. Mater.}\ }\textbf {\bibinfo {volume} {25}},\ \bibinfo
  {pages} {6326} (\bibinfo {year} {2015})}\BibitemShut {NoStop}%
\bibitem [{\citenamefont {Chan}\ \emph {et~al.}(1981)\citenamefont {Chan},
  \citenamefont {Sharma},\ and\ \citenamefont {Smyth}}]{CSS81b}%
  \BibitemOpen
  \bibfield  {author} {\bibinfo {author} {\bibfnamefont {N.~H.}\ \bibnamefont
  {Chan}}, \bibinfo {author} {\bibfnamefont {R.~K.}\ \bibnamefont {Sharma}}, \
  and\ \bibinfo {author} {\bibfnamefont {D.~M.}\ \bibnamefont {Smyth}},\
  }\bibfield  {title} {\emph {\enquote {\bibinfo {title} {{Nonstoichiometry in
  undoped BaTiO$_3$}},}\ }}\href {\doibase 10.1111/j.1151-2916.1981.tb10325.x}
  {\bibfield  {journal} {\bibinfo  {journal} {J. Am. Ceram. Soc.}\ }\textbf
  {\bibinfo {volume} {64}},\ \bibinfo {pages} {556} (\bibinfo {year}
  {1981})}\BibitemShut {NoStop}%
\bibitem [{\citenamefont {Cordero}(2007)}]{Cor07}%
  \BibitemOpen
  \bibfield  {author} {\bibinfo {author} {\bibfnamefont {F.}~\bibnamefont
  {Cordero}},\ }\bibfield  {title} {\emph {\enquote {\bibinfo {title} {{Hopping
  and clustering of oxygen vacancies in SrTiO$_3$ by anelastic relaxation}},}\
  }}\href {\doibase 10.1103/PhysRevB.76.172106} {\bibfield  {journal} {\bibinfo
   {journal} {Phys. Rev. B}\ }\textbf {\bibinfo {volume} {76}},\ \bibinfo
  {pages} {172106} (\bibinfo {year} {2007})}\BibitemShut {NoStop}%
\bibitem [{\citenamefont {Eom}\ \emph {et~al.}(2017)\citenamefont {Eom},
  \citenamefont {Choi}, \citenamefont {Choi}, \citenamefont {Han},
  \citenamefont {Zhou},\ and\ \citenamefont {Lee}}]{ECC17}%
  \BibitemOpen
  \bibfield  {author} {\bibinfo {author} {\bibfnamefont {K.}~\bibnamefont
  {Eom}}, \bibinfo {author} {\bibfnamefont {E.}~\bibnamefont {Choi}}, \bibinfo
  {author} {\bibfnamefont {M.}~\bibnamefont {Choi}}, \bibinfo {author}
  {\bibfnamefont {S.}~\bibnamefont {Han}}, \bibinfo {author} {\bibfnamefont
  {H.}~\bibnamefont {Zhou}}, \ and\ \bibinfo {author} {\bibfnamefont
  {J.}~\bibnamefont {Lee}},\ }\bibfield  {title} {\emph {\enquote {\bibinfo
  {title} {{Oxygen Vacancy Linear Clustering in a Perovskite Oxide}},}\ }}\href
  {\doibase 10.1021/acs.jpclett.7b01348} {\bibfield  {journal} {\bibinfo
  {journal} {J. Phys. Chem. Lett.}\ }\textbf {\bibinfo {volume} {8}},\ \bibinfo
  {pages} {3500} (\bibinfo {year} {2017})}\BibitemShut {NoStop}%
\bibitem [{\citenamefont {Hackmann}\ and\ \citenamefont {Kanert}(1991)}]{HK91}%
  \BibitemOpen
  \bibfield  {author} {\bibinfo {author} {\bibfnamefont {A.}~\bibnamefont
  {Hackmann}}\ and\ \bibinfo {author} {\bibfnamefont {O.}~\bibnamefont
  {Kanert}},\ }\bibfield  {title} {\emph {\enquote {\bibinfo {title} {{NMR
  investigation of defect properties in single crystal SrTiO$_3$}},}\ }}\href
  {\doibase 10.1080/10420159108220797} {\bibfield  {journal} {\bibinfo
  {journal} {Radiation Effects and Defects in Solids}\ }\textbf {\bibinfo
  {volume} {119}},\ \bibinfo {pages} {651} (\bibinfo {year}
  {1991})}\BibitemShut {NoStop}%
\bibitem [{\citenamefont {Buzlukov}\ \emph {et~al.}(2011)\citenamefont
  {Buzlukov}, \citenamefont {Trokiner}, \citenamefont {Kozhevnikov},
  \citenamefont {Verkhovskii}, \citenamefont {Yakubovsky}, \citenamefont
  {Leonidov}, \citenamefont {Gerashenko}, \citenamefont {Stepanov},\ and\
  \citenamefont {Tankeyev}}]{BTK11}%
  \BibitemOpen
  \bibfield  {author} {\bibinfo {author} {\bibfnamefont {A.}~\bibnamefont
  {Buzlukov}}, \bibinfo {author} {\bibfnamefont {A.}~\bibnamefont {Trokiner}},
  \bibinfo {author} {\bibfnamefont {V.}~\bibnamefont {Kozhevnikov}}, \bibinfo
  {author} {\bibfnamefont {S.}~\bibnamefont {Verkhovskii}}, \bibinfo {author}
  {\bibfnamefont {A.}~\bibnamefont {Yakubovsky}}, \bibinfo {author}
  {\bibfnamefont {I.}~\bibnamefont {Leonidov}}, \bibinfo {author}
  {\bibfnamefont {A.}~\bibnamefont {Gerashenko}}, \bibinfo {author}
  {\bibfnamefont {A.}~\bibnamefont {Stepanov}}, \ and\ \bibinfo {author}
  {\bibfnamefont {I.~B.~A.}\ \bibnamefont {Tankeyev}},\ }\bibfield  {title}
  {\emph {\enquote {\bibinfo {title} {{Vacancy ordering and oxygen dynamics in
  oxide ion conducting La$_{1-x}$Sr$_{x}$Ga$_{1-x}$MgxO$_{3-x}$ ceramics:
  $^{71}$Ga, $^{25}$Mg and $^{17}$O NMR}},}\ }}\href {\doibase
  10.1016/j.jssc.2010.10.029} {\bibfield  {journal} {\bibinfo  {journal} {J.
  Solid State Chem.}\ }\textbf {\bibinfo {volume} {184}},\ \bibinfo {pages}
  {36} (\bibinfo {year} {2011})}\BibitemShut {NoStop}%
\bibitem [{\citenamefont {Eichel}(2007)}]{Eic07}%
  \BibitemOpen
  \bibfield  {author} {\bibinfo {author} {\bibfnamefont {R.~A.}\ \bibnamefont
  {Eichel}},\ }\bibfield  {title} {\emph {\enquote {\bibinfo {title} {{Defect
  structure of oxide ferroelectrics-valence state, site of incorporation,
  mechanisms of charge compensation and internal bias fields}},}\ }}\href
  {\doibase 10.1007/s10832-007-9068-8} {\bibfield  {journal} {\bibinfo
  {journal} {J. Electroceram.}\ }\textbf {\bibinfo {volume} {19}},\ \bibinfo
  {pages} {9} (\bibinfo {year} {2007})}\BibitemShut {NoStop}%
\bibitem [{\citenamefont {Tyunina}\ and\ \citenamefont {Savinov}(2020)}]{TS20}%
  \BibitemOpen
  \bibfield  {author} {\bibinfo {author} {\bibfnamefont {M.}~\bibnamefont
  {Tyunina}}\ and\ \bibinfo {author} {\bibfnamefont {M.}~\bibnamefont
  {Savinov}},\ }\bibfield  {title} {\emph {\enquote {\bibinfo {title} {{Charge
  transport in epitaxial barium titanate films}},}\ }}\href {\doibase
  10.1103/PhysRevB.101.094106} {\bibfield  {journal} {\bibinfo  {journal}
  {Phys. Rev. B}\ }\textbf {\bibinfo {volume} {101}},\ \bibinfo {pages}
  {094106} (\bibinfo {year} {2020})}\BibitemShut {NoStop}%
\bibitem [{\citenamefont {Cordero}\ \emph {et~al.}(2019)\citenamefont
  {Cordero}, \citenamefont {Trequattrini}, \citenamefont {Craciun},
  \citenamefont {Langhammer}, \citenamefont {Quiroga},\ and\ \citenamefont
  {P.~S.~Silva}}]{CTC19}%
  \BibitemOpen
  \bibfield  {author} {\bibinfo {author} {\bibfnamefont {F.}~\bibnamefont
  {Cordero}}, \bibinfo {author} {\bibfnamefont {F.}~\bibnamefont
  {Trequattrini}}, \bibinfo {author} {\bibfnamefont {F.}~\bibnamefont
  {Craciun}}, \bibinfo {author} {\bibfnamefont {H.~T.}\ \bibnamefont
  {Langhammer}}, \bibinfo {author} {\bibfnamefont {D.~A.~B.}\ \bibnamefont
  {Quiroga}}, \ and\ \bibinfo {author} {\bibfnamefont {J.}~\bibnamefont
  {P.~S.~Silva}},\ }\bibfield  {title} {\emph {\enquote {\bibinfo {title}
  {{Probing ferroelectricity in highly conducting materials through their
  elastic response: Persistence of ferroelectricity in metallic
  BaTiO$_{3-\delta}$}},}\ }}\href {\doibase 10.1103/PhysRevB.99.064106}
  {\bibfield  {journal} {\bibinfo  {journal} {Phys. Rev. B}\ }\textbf {\bibinfo
  {volume} {99}},\ \bibinfo {pages} {064106} (\bibinfo {year}
  {2019})}\BibitemShut {NoStop}%
\bibitem [{\citenamefont {Cordero}\ \emph {et~al.}(2021)\citenamefont
  {Cordero}, \citenamefont {Trequattrini}, \citenamefont {Quiroga},\ and\
  \citenamefont {Jr.}}]{CTQ21}%
  \BibitemOpen
  \bibfield  {author} {\bibinfo {author} {\bibfnamefont {F.}~\bibnamefont
  {Cordero}}, \bibinfo {author} {\bibfnamefont {F.}~\bibnamefont
  {Trequattrini}}, \bibinfo {author} {\bibfnamefont {D.~A.~B.}\ \bibnamefont
  {Quiroga}}, \ and\ \bibinfo {author} {\bibfnamefont {P.~S.~S.}\ \bibnamefont
  {Jr.}},\ }\bibfield  {title} {\emph {\enquote {\bibinfo {title} {{Hopping and
  clustering of oxygen vacancies in BaTiO$_{3-\delta}$ and the influence of the
  off-centred Ti atoms}},}\ }}\href {\doibase 10.1016/j.jallcom.2021.159753}
  {\bibfield  {journal} {\bibinfo  {journal} {J. Alloys Compd.}\ }\textbf
  {\bibinfo {volume} {874}},\ \bibinfo {pages} {159753} (\bibinfo {year}
  {2021})}\BibitemShut {NoStop}%
\bibitem [{\citenamefont {Cordero}\ \emph {et~al.}(2023)\citenamefont
  {Cordero}, \citenamefont {Trequattrini}, \citenamefont {da~Silva},
  \citenamefont {Jr.}, \citenamefont {Venet}, \citenamefont {Aktas},\ and\
  \citenamefont {Salje}}]{CTS23}%
  \BibitemOpen
  \bibfield  {author} {\bibinfo {author} {\bibfnamefont {F.}~\bibnamefont
  {Cordero}}, \bibinfo {author} {\bibfnamefont {F.}~\bibnamefont
  {Trequattrini}}, \bibinfo {author} {\bibfnamefont {P.~S.}\ \bibnamefont
  {da~Silva}}, \bibinfo {author} {\bibnamefont {Jr.}}, \bibinfo {author}
  {\bibfnamefont {M.}~\bibnamefont {Venet}}, \bibinfo {author} {\bibfnamefont
  {O.}~\bibnamefont {Aktas}}, \ and\ \bibinfo {author} {\bibfnamefont
  {E.~K.~H.}\ \bibnamefont {Salje}},\ }\bibfield  {title} {\emph {\enquote
  {\bibinfo {title} {{Elastic precursor effects during Ba$_{1-x}$Sr$_x$TiO$_3$
  ferroelastic phase transitions}},}\ }}\href {\doibase
  10.1103/PhysRevResearch.5.013121} {\bibfield  {journal} {\bibinfo  {journal}
  {Phys. Rev. Research}\ }\textbf {\bibinfo {volume} {5}},\ \bibinfo {pages}
  {013121} (\bibinfo {year} {2023})}\BibitemShut {NoStop}%
\bibitem [{\citenamefont {Cordero}\ \emph {et~al.}(2009)\citenamefont
  {Cordero}, \citenamefont {{Dalla Bella}}, \citenamefont {Corvasce},
  \citenamefont {Latino},\ and\ \citenamefont {Morbidini}}]{CDC09}%
  \BibitemOpen
  \bibfield  {author} {\bibinfo {author} {\bibfnamefont {F.}~\bibnamefont
  {Cordero}}, \bibinfo {author} {\bibfnamefont {L.}~\bibnamefont {{Dalla
  Bella}}}, \bibinfo {author} {\bibfnamefont {F.}~\bibnamefont {Corvasce}},
  \bibinfo {author} {\bibfnamefont {P.~M.}\ \bibnamefont {Latino}}, \ and\
  \bibinfo {author} {\bibfnamefont {A.}~\bibnamefont {Morbidini}},\ }\bibfield
  {title} {\emph {\enquote {\bibinfo {title} {{An insert for anelastic
  spectroscopy measurements from 80~K to 1100~K}},}\ }}\href {\doibase
  10.1088/0957-0233/20/1/015702} {\bibfield  {journal} {\bibinfo  {journal}
  {Meas. Sci. Technol.}\ }\textbf {\bibinfo {volume} {20}},\ \bibinfo {pages}
  {015702} (\bibinfo {year} {2009})}\BibitemShut {NoStop}%
\bibitem [{\citenamefont {Nowick}\ and\ \citenamefont {Berry}(1972)}]{NB72}%
  \BibitemOpen
  \bibfield  {author} {\bibinfo {author} {\bibfnamefont {A.~S.}\ \bibnamefont
  {Nowick}}\ and\ \bibinfo {author} {\bibfnamefont {B.~S.}\ \bibnamefont
  {Berry}},\ }\href@noop {} {\emph {\bibinfo {title} {{Anelastic Relaxation in
  Crystalline Solids}}}}\ (\bibinfo  {publisher} {Academic Press},\ \bibinfo
  {address} {New York},\ \bibinfo {year} {1972})\BibitemShut {NoStop}%
\bibitem [{\citenamefont {Cordero}(1993)}]{Cor93}%
  \BibitemOpen
  \bibfield  {author} {\bibinfo {author} {\bibfnamefont {F.}~\bibnamefont
  {Cordero}},\ }\bibfield  {title} {\emph {\enquote {\bibinfo {title}
  {{Anelastic (dielectric) relaxation of point defects at any concentration,
  with blocking effects and formation of complexes}},}\ }}\href {\doibase
  10.1103/PhysRevB.47.7674} {\bibfield  {journal} {\bibinfo  {journal} {Phys.
  Rev. B}\ }\textbf {\bibinfo {volume} {47}},\ \bibinfo {pages} {7674}
  (\bibinfo {year} {1993})}\BibitemShut {NoStop}%
\bibitem [{\citenamefont {Cannelli}\ \emph {et~al.}(1994)\citenamefont
  {Cannelli}, \citenamefont {Cantelli}, \citenamefont {Cordero}, \citenamefont
  {Piraccini}, \citenamefont {Trequattrini},\ and\ \citenamefont
  {Ferretti}}]{44}%
  \BibitemOpen
  \bibfield  {author} {\bibinfo {author} {\bibfnamefont {G.}~\bibnamefont
  {Cannelli}}, \bibinfo {author} {\bibfnamefont {R.}~\bibnamefont {Cantelli}},
  \bibinfo {author} {\bibfnamefont {F.}~\bibnamefont {Cordero}}, \bibinfo
  {author} {\bibfnamefont {N.}~\bibnamefont {Piraccini}}, \bibinfo {author}
  {\bibfnamefont {F.}~\bibnamefont {Trequattrini}}, \ and\ \bibinfo {author}
  {\bibfnamefont {M.}~\bibnamefont {Ferretti}},\ }\bibfield  {title} {\emph
  {\enquote {\bibinfo {title} {{Mobility and aggregation of oxygen in
  YBa$_2$Cu$_3$O$_{6+x}$ in the low- concentration limit}},}\ }}\href {\doibase
  10.1103/PhysRevB.50.16679} {\bibfield  {journal} {\bibinfo  {journal} {Phys.
  Rev. B}\ }\textbf {\bibinfo {volume} {50}},\ \bibinfo {pages} {16679}
  (\bibinfo {year} {1994})}\BibitemShut {NoStop}%
\bibitem [{\citenamefont {Chandrasekaran}\ \emph {et~al.}(2016)\citenamefont
  {Chandrasekaran}, \citenamefont {Wei}, \citenamefont {Feigl}, \citenamefont
  {Damjanovic}, \citenamefont {Setter},\ and\ \citenamefont
  {Marzari}}]{CWF16b}%
  \BibitemOpen
  \bibfield  {author} {\bibinfo {author} {\bibfnamefont {A.}~\bibnamefont
  {Chandrasekaran}}, \bibinfo {author} {\bibfnamefont {X.~K.}\ \bibnamefont
  {Wei}}, \bibinfo {author} {\bibfnamefont {L.}~\bibnamefont {Feigl}}, \bibinfo
  {author} {\bibfnamefont {D.}~\bibnamefont {Damjanovic}}, \bibinfo {author}
  {\bibfnamefont {N.}~\bibnamefont {Setter}}, \ and\ \bibinfo {author}
  {\bibfnamefont {N.}~\bibnamefont {Marzari}},\ }\bibfield  {title} {\emph
  {\enquote {\bibinfo {title} {{Asymmetric structure of 90$^{\circ}$ domain
  walls and interactions with defects in PbTiO$_3$}},}\ }}\href {\doibase
  10.1103/PhysRevB.93.144102} {\bibfield  {journal} {\bibinfo  {journal} {Phys.
  Rev. B}\ }\textbf {\bibinfo {volume} {93}},\ \bibinfo {pages} {144102}
  (\bibinfo {year} {2016})}\BibitemShut {NoStop}%
\bibitem [{\citenamefont {Sun}\ \emph {et~al.}(2005)\citenamefont {Sun},
  \citenamefont {Ren},\ and\ \citenamefont {Otsuka}}]{SRO05b}%
  \BibitemOpen
  \bibfield  {author} {\bibinfo {author} {\bibfnamefont {D.}~\bibnamefont
  {Sun}}, \bibinfo {author} {\bibfnamefont {X.}~\bibnamefont {Ren}}, \ and\
  \bibinfo {author} {\bibfnamefont {K.}~\bibnamefont {Otsuka}},\ }\bibfield
  {title} {\emph {\enquote {\bibinfo {title} {{Stabilization effect in
  ferroelectric materials during aging in ferroelectric state}},}\ }}\href
  {\doibase 10.1063/1.2084343} {\bibfield  {journal} {\bibinfo  {journal}
  {Appl. Phys. Lett.}\ }\textbf {\bibinfo {volume} {87}},\ \bibinfo {pages}
  {142903} (\bibinfo {year} {2005})}\BibitemShut {NoStop}%
\bibitem [{\citenamefont {Carl}\ and\ \citenamefont {Hardtl}(1977)}]{CH77}%
  \BibitemOpen
  \bibfield  {author} {\bibinfo {author} {\bibfnamefont {K.}~\bibnamefont
  {Carl}}\ and\ \bibinfo {author} {\bibfnamefont {K.~H.}\ \bibnamefont
  {Hardtl}},\ }\bibfield  {title} {\emph {\enquote {\bibinfo {title}
  {{Electrical after-effects in Pb(Ti, Zr)O$_3$ceramics}},}\ }}\href {\doibase
  10.1080/00150197808236770} {\bibfield  {journal} {\bibinfo  {journal}
  {Ferroelectrics}\ }\textbf {\bibinfo {volume} {17}},\ \bibinfo {pages} {473}
  (\bibinfo {year} {1977})}\BibitemShut {NoStop}%
\bibitem [{\citenamefont {Ren}(2004)}]{Ren04}%
  \BibitemOpen
  \bibfield  {author} {\bibinfo {author} {\bibfnamefont {X.}~\bibnamefont
  {Ren}},\ }\bibfield  {title} {\emph {\enquote {\bibinfo {title} {Large
  electric-field-induced strain in ferroelectric crystals by
  point-defect-mediated reversible domain switching},}\ }}\href {\doibase
  10.1038/nmat1051} {\bibfield  {journal} {\bibinfo  {journal} {Nat. Mater.}\
  }\textbf {\bibinfo {volume} {3}},\ \bibinfo {pages} {91} (\bibinfo {year}
  {2004})}\BibitemShut {NoStop}%
\bibitem [{\citenamefont {Granhed}\ \emph {et~al.}(2019)\citenamefont
  {Granhed}, \citenamefont {Lindman}, \citenamefont {Ekl{\"o}f-{\"O}sterberg},
  \citenamefont {Karlsson}, \citenamefont {Parker},\ and\ \citenamefont
  {Wahnstr{\"o}m}}]{GLE19}%
  \BibitemOpen
  \bibfield  {author} {\bibinfo {author} {\bibfnamefont {E.~J.}\ \bibnamefont
  {Granhed}}, \bibinfo {author} {\bibfnamefont {A.}~\bibnamefont {Lindman}},
  \bibinfo {author} {\bibfnamefont {C.}~\bibnamefont
  {Ekl{\"o}f-{\"O}sterberg}}, \bibinfo {author} {\bibfnamefont
  {M.}~\bibnamefont {Karlsson}}, \bibinfo {author} {\bibfnamefont {S.~F.}\
  \bibnamefont {Parker}}, \ and\ \bibinfo {author} {\bibfnamefont
  {G.}~\bibnamefont {Wahnstr{\"o}m}},\ }\bibfield  {title} {\emph {\enquote
  {\bibinfo {title} {{Band vs. polaron: vibrational motion and chemical
  expansion of hydride ions as signatures for the electronic character in
  oxyhydride barium titanate}},}\ }}\href {\doibase 10.1039/c9ta00086k}
  {\bibfield  {journal} {\bibinfo  {journal} {J. Mater. Chem. A}\ }\textbf
  {\bibinfo {volume} {7}},\ \bibinfo {pages} {16211} (\bibinfo {year}
  {2019})}\BibitemShut {NoStop}%
\bibitem [{\citenamefont {Nossa}\ \emph {et~al.}(2015)\citenamefont {Nossa},
  \citenamefont {Naumov},\ and\ \citenamefont {Cohen}}]{NNC15}%
  \BibitemOpen
  \bibfield  {author} {\bibinfo {author} {\bibfnamefont {J.~F.}\ \bibnamefont
  {Nossa}}, \bibinfo {author} {\bibfnamefont {I.~I.}\ \bibnamefont {Naumov}}, \
  and\ \bibinfo {author} {\bibfnamefont {R.~E.}\ \bibnamefont {Cohen}},\
  }\bibfield  {title} {\emph {\enquote {\bibinfo {title} {{Effects of manganese
  addition on the electronic structure of BaTiO$_3$}},}\ }}\href {\doibase
  10.1103/PhysRevB.91.214105} {\bibfield  {journal} {\bibinfo  {journal} {Phys.
  Rev. B}\ }\textbf {\bibinfo {volume} {91}},\ \bibinfo {pages} {214105}
  (\bibinfo {year} {2015})}\BibitemShut {NoStop}%
\bibitem [{\citenamefont {Shimada}\ \emph {et~al.}(2013)\citenamefont
  {Shimada}, \citenamefont {Ueda}, \citenamefont {Wang},\ and\ \citenamefont
  {Kitamura}}]{SUW13}%
  \BibitemOpen
  \bibfield  {author} {\bibinfo {author} {\bibfnamefont {T.}~\bibnamefont
  {Shimada}}, \bibinfo {author} {\bibfnamefont {T.}~\bibnamefont {Ueda}},
  \bibinfo {author} {\bibfnamefont {J.}~\bibnamefont {Wang}}, \ and\ \bibinfo
  {author} {\bibfnamefont {T.}~\bibnamefont {Kitamura}},\ }\bibfield  {title}
  {\emph {\enquote {\bibinfo {title} {{Hybrid Hartree-Fock density functional
  study of charged point defects in ferroelectric PbTiO$_3$}},}\ }}\href
  {\doibase 10.1103/PhysRevB.87.174111} {\bibfield  {journal} {\bibinfo
  {journal} {Phys. Rev. B}\ }\textbf {\bibinfo {volume} {87}},\ \bibinfo
  {pages} {174111} (\bibinfo {year} {2013})}\BibitemShut {NoStop}%
\bibitem [{\citenamefont {Yao}\ and\ \citenamefont {Fu}(2011)}]{YF11}%
  \BibitemOpen
  \bibfield  {author} {\bibinfo {author} {\bibfnamefont {Y.}~\bibnamefont
  {Yao}}\ and\ \bibinfo {author} {\bibfnamefont {H.}~\bibnamefont {Fu}},\
  }\bibfield  {title} {\emph {\enquote {\bibinfo {title} {{Charged vacancies in
  ferroelectric PbTiO$_3$: Formation energies, optimal Fermi region, and
  influence on local polarization}},}\ }}\href {\doibase
  10.1103/PhysRevB.84.064112} {\bibfield  {journal} {\bibinfo  {journal} {Phys.
  Rev. B}\ }\textbf {\bibinfo {volume} {84}},\ \bibinfo {pages} {064112}
  (\bibinfo {year} {2011})}\BibitemShut {NoStop}%
\bibitem [{\citenamefont {Xu}\ \emph {et~al.}(2016)\citenamefont {Xu},
  \citenamefont {Shimada}, \citenamefont {Araki}, \citenamefont {Wang},\ and\
  \citenamefont {Kitamura}}]{XSA16}%
  \BibitemOpen
  \bibfield  {author} {\bibinfo {author} {\bibfnamefont {T.}~\bibnamefont
  {Xu}}, \bibinfo {author} {\bibfnamefont {T.}~\bibnamefont {Shimada}},
  \bibinfo {author} {\bibfnamefont {Y.}~\bibnamefont {Araki}}, \bibinfo
  {author} {\bibfnamefont {J.}~\bibnamefont {Wang}}, \ and\ \bibinfo {author}
  {\bibfnamefont {T.}~\bibnamefont {Kitamura}},\ }\bibfield  {title} {\emph
  {\enquote {\bibinfo {title} {{Multiferroic Domain Walls in Ferroelectric
  PbTiO$_3$ with Oxygen Deficiency}},}\ }}\href {\doibase
  10.1021/acs.nanolett.5b04113} {\bibfield  {journal} {\bibinfo  {journal}
  {Nano Lett.}\ }\textbf {\bibinfo {volume} {16}},\ \bibinfo {pages} {454}
  (\bibinfo {year} {2016})}\BibitemShut {NoStop}%
\bibitem [{\citenamefont {Wang}\ \emph {et~al.}(2019)\citenamefont {Wang},
  \citenamefont {Xu}, \citenamefont {Xuan}, \citenamefont {Chen}, \citenamefont
  {Shimada},\ and\ \citenamefont {Kitamura}}]{WXX19}%
  \BibitemOpen
  \bibfield  {author} {\bibinfo {author} {\bibfnamefont {X.}~\bibnamefont
  {Wang}}, \bibinfo {author} {\bibfnamefont {T.}~\bibnamefont {Xu}}, \bibinfo
  {author} {\bibfnamefont {F.}~\bibnamefont {Xuan}}, \bibinfo {author}
  {\bibfnamefont {C.}~\bibnamefont {Chen}}, \bibinfo {author} {\bibfnamefont
  {T.}~\bibnamefont {Shimada}}, \ and\ \bibinfo {author} {\bibfnamefont
  {T.}~\bibnamefont {Kitamura}},\ }\bibfield  {title} {\emph {\enquote
  {\bibinfo {title} {{Effect of the oxygen vacancy on the ferroelectricity of
  90$^{\circ}$ domain wall structure in PbTiO$_3$: A density functional theory
  study}},}\ }}\href {\doibase 10.1063/1.5125306} {\bibfield  {journal}
  {\bibinfo  {journal} {J. Appl. Phys.}\ }\textbf {\bibinfo {volume} {126}},\
  \bibinfo {pages} {174107} (\bibinfo {year} {2019})}\BibitemShut {NoStop}%
\bibitem [{\citenamefont {Alahmed}\ and\ \citenamefont {Fu}(2007)}]{AF07}%
  \BibitemOpen
  \bibfield  {author} {\bibinfo {author} {\bibfnamefont {Z.}~\bibnamefont
  {Alahmed}}\ and\ \bibinfo {author} {\bibfnamefont {H.~X.}\ \bibnamefont
  {Fu}},\ }\bibfield  {title} {\emph {\enquote {\bibinfo {title}
  {{First-principles determination of chemical potentials and vacancy formation
  energies in PbTiO$_3$ and BaTiO$_3$}},}\ }}\href {\doibase
  10.1103/PhysRevB.76.224101} {\bibfield  {journal} {\bibinfo  {journal} {Phys.
  Rev. B}\ }\textbf {\bibinfo {volume} {76}},\ \bibinfo {pages} {224101}
  (\bibinfo {year} {2007})}\BibitemShut {NoStop}%
\bibitem [{\citenamefont {Ahmad}\ \emph {et~al.}(2007)\citenamefont {Ahmad},
  \citenamefont {Yamada}, \citenamefont {Meuffels},\ and\ \citenamefont
  {Waser}}]{AYM07}%
  \BibitemOpen
  \bibfield  {author} {\bibinfo {author} {\bibfnamefont {M.~M.}\ \bibnamefont
  {Ahmad}}, \bibinfo {author} {\bibfnamefont {K.}~\bibnamefont {Yamada}},
  \bibinfo {author} {\bibfnamefont {P.}~\bibnamefont {Meuffels}}, \ and\
  \bibinfo {author} {\bibfnamefont {R.}~\bibnamefont {Waser}},\ }\bibfield
  {title} {\emph {\enquote {\bibinfo {title} {Aging-induced dielectric
  relaxation in barium titanate ceramics},}\ }}\href {\doibase
  10.1063/1.2713178} {\bibfield  {journal} {\bibinfo  {journal} {Appl. Phys.
  Lett.}\ }\textbf {\bibinfo {volume} {90}},\ \bibinfo {pages} {112902}
  (\bibinfo {year} {2007})}\BibitemShut {NoStop}%
\bibitem [{\citenamefont {Lee}\ \emph {et~al.}(2007)\citenamefont {Lee},
  \citenamefont {Liu}, \citenamefont {Kim},\ and\ \citenamefont
  {Randall}}]{LLK07}%
  \BibitemOpen
  \bibfield  {author} {\bibinfo {author} {\bibfnamefont {S.}~\bibnamefont
  {Lee}}, \bibinfo {author} {\bibfnamefont {Z.-K.}\ \bibnamefont {Liu}},
  \bibinfo {author} {\bibfnamefont {M.-H.}\ \bibnamefont {Kim}}, \ and\
  \bibinfo {author} {\bibfnamefont {C.~A.}\ \bibnamefont {Randall}},\
  }\bibfield  {title} {\emph {\enquote {\bibinfo {title} {{Influence of
  nonstoichiometry on ferroelectric phase transition in BaTiO$_{3}$}},}\
  }}\href {\doibase 10.1063/1.2710280} {\bibfield  {journal} {\bibinfo
  {journal} {J. Appl. Phys.}\ }\textbf {\bibinfo {volume} {101}},\ \bibinfo
  {pages} {054119} (\bibinfo {year} {2007})}\BibitemShut {NoStop}%
\bibitem [{\citenamefont {Lemanov}\ \emph {et~al.}(1996)\citenamefont
  {Lemanov}, \citenamefont {Smirnova}, \citenamefont {Syrnikov},\ and\
  \citenamefont {Tarakanov}}]{LSS96}%
  \BibitemOpen
  \bibfield  {author} {\bibinfo {author} {\bibfnamefont {V.~V.}\ \bibnamefont
  {Lemanov}}, \bibinfo {author} {\bibfnamefont {E.~P.}\ \bibnamefont
  {Smirnova}}, \bibinfo {author} {\bibfnamefont {P.~P.}\ \bibnamefont
  {Syrnikov}}, \ and\ \bibinfo {author} {\bibfnamefont {E.~A.}\ \bibnamefont
  {Tarakanov}},\ }\bibfield  {title} {\emph {\enquote {\bibinfo {title} {{Phase
  transitions and glasslike behavior in Sr$_{1-x}$Ba$_x$TiO$_3$}},}\ }}\href
  {\doibase 10.1103/PhysRevB.54.3151} {\bibfield  {journal} {\bibinfo
  {journal} {Phys. Rev. B}\ }\textbf {\bibinfo {volume} {54}},\ \bibinfo
  {pages} {3151} (\bibinfo {year} {1996})}\BibitemShut {NoStop}%
\bibitem [{\citenamefont {Cordero}\ \emph {et~al.}(2016)\citenamefont
  {Cordero}, \citenamefont {Craciun}, \citenamefont {Trequattrini},\ and\
  \citenamefont {Galassi}}]{CCT16}%
  \BibitemOpen
  \bibfield  {author} {\bibinfo {author} {\bibfnamefont {F.}~\bibnamefont
  {Cordero}}, \bibinfo {author} {\bibfnamefont {F.}~\bibnamefont {Craciun}},
  \bibinfo {author} {\bibfnamefont {F.}~\bibnamefont {Trequattrini}}, \ and\
  \bibinfo {author} {\bibfnamefont {C.}~\bibnamefont {Galassi}},\ }\bibfield
  {title} {\emph {\enquote {\bibinfo {title} {{Piezoelectric softening in
  ferroelectrics: ferroelectric versus antiferroelectric
  PbZr$_{1-x}$Ti$_{x}$O$_{3}$}},}\ }}\href {\doibase
  10.1103/PhysRevB.93.174111} {\bibfield  {journal} {\bibinfo  {journal} {Phys.
  Rev. B}\ }\textbf {\bibinfo {volume} {93}},\ \bibinfo {pages} {174111}
  (\bibinfo {year} {2016})}\BibitemShut {NoStop}%
\bibitem [{\citenamefont {Ondrejkovic}\ \emph {et~al.}(2013)\citenamefont
  {Ondrejkovic}, \citenamefont {Marton}, \citenamefont {Guennou}, \citenamefont
  {Setter},\ and\ \citenamefont {Hlinka}}]{OMG13}%
  \BibitemOpen
  \bibfield  {author} {\bibinfo {author} {\bibfnamefont {P.}~\bibnamefont
  {Ondrejkovic}}, \bibinfo {author} {\bibfnamefont {P.}~\bibnamefont {Marton}},
  \bibinfo {author} {\bibfnamefont {M.}~\bibnamefont {Guennou}}, \bibinfo
  {author} {\bibfnamefont {N.}~\bibnamefont {Setter}}, \ and\ \bibinfo {author}
  {\bibfnamefont {J.}~\bibnamefont {Hlinka}},\ }\bibfield  {title} {\emph
  {\enquote {\bibinfo {title} {Piezoelectric properties of twinned
  ferroelectric perovskites with head-to-head and tail-to-tail domain walls},}\
  }}\href {\doibase 10.1103/PhysRevB.88.024114} {\bibfield  {journal} {\bibinfo
   {journal} {Phys. Rev. B}\ }\textbf {\bibinfo {volume} {88}},\ \bibinfo
  {pages} {024114} (\bibinfo {year} {2013})}\BibitemShut {NoStop}%
\bibitem [{\citenamefont {Sluka}\ \emph {et~al.}(2012)\citenamefont {Sluka},
  \citenamefont {Tagantsev}, \citenamefont {Damjanovic}, \citenamefont
  {Gureev},\ and\ \citenamefont {Setter}}]{STD12}%
  \BibitemOpen
  \bibfield  {author} {\bibinfo {author} {\bibfnamefont {T.}~\bibnamefont
  {Sluka}}, \bibinfo {author} {\bibfnamefont {A.~K.}\ \bibnamefont
  {Tagantsev}}, \bibinfo {author} {\bibfnamefont {D.}~\bibnamefont
  {Damjanovic}}, \bibinfo {author} {\bibfnamefont {M.}~\bibnamefont {Gureev}},
  \ and\ \bibinfo {author} {\bibfnamefont {N.}~\bibnamefont {Setter}},\
  }\bibfield  {title} {\emph {\enquote {\bibinfo {title} {Enhanced
  electromechanical response of ferroelectrics due to charged domain walls},}\
  }}\href {\doibase 10.1038/ncomms1751} {\bibfield  {journal} {\bibinfo
  {journal} {Nat. Commun.}\ }\textbf {\bibinfo {volume} {4}},\ \bibinfo {pages}
  {1751} (\bibinfo {year} {2012})}\BibitemShut {NoStop}%
\bibitem [{\citenamefont {Petralanda}\ \emph {et~al.}(2021)\citenamefont
  {Petralanda}, \citenamefont {Kruse}, \citenamefont {Simons},\ and\
  \citenamefont {Olsen}}]{PKS21}%
  \BibitemOpen
  \bibfield  {author} {\bibinfo {author} {\bibfnamefont {U.}~\bibnamefont
  {Petralanda}}, \bibinfo {author} {\bibfnamefont {M.}~\bibnamefont {Kruse}},
  \bibinfo {author} {\bibfnamefont {H.}~\bibnamefont {Simons}}, \ and\ \bibinfo
  {author} {\bibfnamefont {T.}~\bibnamefont {Olsen}},\ }\bibfield  {title}
  {\emph {\enquote {\bibinfo {title} {{Oxygen Vacancies Nucleate Charged Domain
  Walls in Ferroelectrics}},}\ }}\href {\doibase
  10.1103/PhysRevLett.127.117601} {\bibfield  {journal} {\bibinfo  {journal}
  {Phys. Rev. Lett.}\ }\textbf {\bibinfo {volume} {127}},\ \bibinfo {pages}
  {117601} (\bibinfo {year} {2021})}\BibitemShut {NoStop}%
\bibitem [{\citenamefont {Hong}\ \emph {et~al.}(2008)\citenamefont {Hong},
  \citenamefont {Soh}, \citenamefont {Du},\ and\ \citenamefont {Li}}]{HSD08}%
  \BibitemOpen
  \bibfield  {author} {\bibinfo {author} {\bibfnamefont {L.}~\bibnamefont
  {Hong}}, \bibinfo {author} {\bibfnamefont {A.~K.}\ \bibnamefont {Soh}},
  \bibinfo {author} {\bibfnamefont {Q.~G.}\ \bibnamefont {Du}}, \ and\ \bibinfo
  {author} {\bibfnamefont {J.~Y.}\ \bibnamefont {Li}},\ }\bibfield  {title}
  {\emph {\enquote {\bibinfo {title} {Interaction of o vacancies and domain
  structures in single crystal {BaTiO$_3$}: Two-dimensional ferroelectric
  model},}\ }}\href {\doibase 10.1103/PhysRevB.77.094104} {\bibfield  {journal}
  {\bibinfo  {journal} {Phys. Rev. B}\ }\textbf {\bibinfo {volume} {77}},\
  \bibinfo {pages} {094104} (\bibinfo {year} {2008})}\BibitemShut {NoStop}%
\bibitem [{\citenamefont {Chandrasekaran}\ \emph {et~al.}(2013)\citenamefont
  {Chandrasekaran}, \citenamefont {Damjanovic}, \citenamefont {Setter},\ and\
  \citenamefont {Marzari}}]{CDS13}%
  \BibitemOpen
  \bibfield  {author} {\bibinfo {author} {\bibfnamefont {A.}~\bibnamefont
  {Chandrasekaran}}, \bibinfo {author} {\bibfnamefont {D.}~\bibnamefont
  {Damjanovic}}, \bibinfo {author} {\bibfnamefont {N.}~\bibnamefont {Setter}},
  \ and\ \bibinfo {author} {\bibfnamefont {N.}~\bibnamefont {Marzari}},\
  }\bibfield  {title} {\emph {\enquote {\bibinfo {title} {{Defect ordering and
  defect-domain-wall interactions in PbTiO$_3$: A first-principles study}},}\
  }}\href {\doibase 10.1103/PhysRevB.88.214116} {\bibfield  {journal} {\bibinfo
   {journal} {Phys. Rev. B}\ }\textbf {\bibinfo {volume} {88}},\ \bibinfo
  {pages} {214116} (\bibinfo {year} {2013})}\BibitemShut {NoStop}%
\bibitem [{\citenamefont {He}\ and\ \citenamefont {Vanderbilt}(2003)}]{HV03}%
  \BibitemOpen
  \bibfield  {author} {\bibinfo {author} {\bibfnamefont {L.}~\bibnamefont
  {He}}\ and\ \bibinfo {author} {\bibfnamefont {D.}~\bibnamefont
  {Vanderbilt}},\ }\bibfield  {title} {\emph {\enquote {\bibinfo {title}
  {{First-principles study of oxygen-vacancy pinning of domain walls in
  PbTiO$_3$}},}\ }}\href {\doibase 10.1103/PhysRevB.68.134103} {\bibfield
  {journal} {\bibinfo  {journal} {Phys. Rev. B}\ }\textbf {\bibinfo {volume}
  {68}},\ \bibinfo {pages} {134103} (\bibinfo {year} {2003})}\BibitemShut
  {NoStop}%
\bibitem [{\citenamefont {Samanta}\ \emph {et~al.}(2021)\citenamefont
  {Samanta}, \citenamefont {Yadav}, \citenamefont {Gu}, \citenamefont {Meyers},
  \citenamefont {Wu}, \citenamefont {Chen}, \citenamefont {Pandya},
  \citenamefont {York}, \citenamefont {Martin}, \citenamefont {Spanier},\ and\
  \citenamefont {Grinberg}}]{SYG21}%
  \BibitemOpen
  \bibfield  {author} {\bibinfo {author} {\bibfnamefont {A.}~\bibnamefont
  {Samanta}}, \bibinfo {author} {\bibfnamefont {S.}~\bibnamefont {Yadav}},
  \bibinfo {author} {\bibfnamefont {Z.}~\bibnamefont {Gu}}, \bibinfo {author}
  {\bibfnamefont {C.~G.}\ \bibnamefont {Meyers}}, \bibinfo {author}
  {\bibfnamefont {L.}~\bibnamefont {Wu}}, \bibinfo {author} {\bibfnamefont
  {D.}~\bibnamefont {Chen}}, \bibinfo {author} {\bibfnamefont {S.}~\bibnamefont
  {Pandya}}, \bibinfo {author} {\bibfnamefont {R.~A.}\ \bibnamefont {York}},
  \bibinfo {author} {\bibfnamefont {L.~W.}\ \bibnamefont {Martin}}, \bibinfo
  {author} {\bibfnamefont {J.~E.}\ \bibnamefont {Spanier}}, \ and\ \bibinfo
  {author} {\bibfnamefont {I.}~\bibnamefont {Grinberg}},\ }\bibfield  {title}
  {\emph {\enquote {\bibinfo {title} {{A Predictive Theory for Domain Walls in
  Oxide Ferroelectrics Based on Interatomic Interactions and its Implications
  for Collective Material Properties}},}\ }}\href {\doibase
  10.1002/adma.202106021} {\bibfield  {journal} {\bibinfo  {journal} {Adv.
  Mater.}\ }\textbf {\bibinfo {volume} {34}},\ \bibinfo {pages} {2106021}
  (\bibinfo {year} {2021})}\BibitemShut {NoStop}%
\bibitem [{\citenamefont {Fan}\ \emph {et~al.}(2018)\citenamefont {Fan},
  \citenamefont {Koruza}, \citenamefont {R{\"o}del},\ and\ \citenamefont
  {Tan}}]{FKR18}%
  \BibitemOpen
  \bibfield  {author} {\bibinfo {author} {\bibfnamefont {Z.}~\bibnamefont
  {Fan}}, \bibinfo {author} {\bibfnamefont {J.}~\bibnamefont {Koruza}},
  \bibinfo {author} {\bibfnamefont {J.}~\bibnamefont {R{\"o}del}}, \ and\
  \bibinfo {author} {\bibfnamefont {X.}~\bibnamefont {Tan}},\ }\bibfield
  {title} {\emph {\enquote {\bibinfo {title} {{An ideal amplitude window
  against electric fatigue in BaTiO$_3$-based lead-free piezoelectric
  materials}},}\ }}\href {\doibase 10.1016/j.actamat.2018.03.067} {\bibfield
  {journal} {\bibinfo  {journal} {Acta Mater.}\ }\textbf {\bibinfo {volume}
  {151}},\ \bibinfo {pages} {253} (\bibinfo {year} {2018})}\BibitemShut
  {NoStop}%
\bibitem [{\citenamefont {Zhang}\ \emph {et~al.}(2015)\citenamefont {Zhang},
  \citenamefont {Glaum}, \citenamefont {Ehmke}, \citenamefont {Bowman},
  \citenamefont {Blendell},\ and\ \citenamefont {Hoffman}}]{ZGE15}%
  \BibitemOpen
  \bibfield  {author} {\bibinfo {author} {\bibfnamefont {Y.}~\bibnamefont
  {Zhang}}, \bibinfo {author} {\bibfnamefont {J.}~\bibnamefont {Glaum}},
  \bibinfo {author} {\bibfnamefont {M.~C.}\ \bibnamefont {Ehmke}}, \bibinfo
  {author} {\bibfnamefont {K.~J.}\ \bibnamefont {Bowman}}, \bibinfo {author}
  {\bibfnamefont {J.~E.}\ \bibnamefont {Blendell}}, \ and\ \bibinfo {author}
  {\bibfnamefont {M.~J.}\ \bibnamefont {Hoffman}},\ }\bibfield  {title} {\emph
  {\enquote {\bibinfo {title} {{The ageing and de-ageing behaviour of
  (Ba$_{0.8}$5Ca$_{0.15}$)(Ti$_{0.9}$Zr$_{0.1})$O$_3$ lead-free piezoelectric
  ceramics}},}\ }}\href {\doibase 10.1063/1.4931892} {\bibfield  {journal}
  {\bibinfo  {journal} {J. Appl. Phys.}\ }\textbf {\bibinfo {volume} {118}},\
  \bibinfo {pages} {124108} (\bibinfo {year} {2015})}\BibitemShut {NoStop}%
\end{thebibliography}%

\end{document}